\documentclass[draft]{agujournal2019}

\usepackage{changes}
\usepackage{xifthen}

\newboolean{showchanges}
\setboolean{showchanges}{false}  
\ifthenelse{\boolean{showchanges}}{
}{%
    
    \renewcommand{\deleted}[1]{}
    
}

\usepackage{url} 
\usepackage{lineno}
\usepackage[inline]{trackchanges} 
\usepackage{soul}
\usepackage{amsmath}
\usepackage{gensymb}
\usepackage{array}
\usepackage{bbold}
\usepackage{natbib}

\usepackage{hyperref}

\draftfalse

\journalname{Journal of Advances in Modeling Earth Systems (JAMES)}

\begin{document}


\title{Prediction Beyond the Medium Range with an Atmosphere-Ocean Model that Combines Physics-based Modeling and Machine Learning}

\authors{Dhruvit Patel\affil{1}\thanks{Currently at the University of Chicago}, Troy Arcomano\affil{2}, Brian Hunt\affil{3}, Istvan Szunyogh\affil{4}, and Edward Ott\affil{1,5}}

\affiliation{1}{Department of Physics, University of Maryland, College Park, MD, USA}
\affiliation{2}{Argonne National Laboratory, Lemont, IL, USA}
\affiliation{3}{Institute for Physical Science and Technology, University of Maryland, College Park, MD, USA}
\affiliation{4}{Department of Atmospheric Sciences, Texas A\&M University, College Station, TX, USA}
\affiliation{5}{Department of Electrical and Computer Engineering, University of Maryland, College Park, MD, USA}

\correspondingauthor{Dhruvit Patel}{dpp94@uchicago.edu}


\begin{keypoints}
\item A low-resolution atmospheric general circulation model hybridized with machine learning has forecast skill beyond the medium range.
\item The low-resolution hybrid model has prediction skill comparable to that of high-resolution, purely physics-based models for many tasks.
\item Compared to a conventional high-resolution purely physics-based approach, our hybrid approach requires much less computational resources.
\end{keypoints}


\begin{abstract}
This paper explores the potential of a hybrid modeling approach that combines machine learning (ML) with conventional physics-based modeling for weather prediction beyond the medium range. It extends the work of \cite{Arcomano_2022}, which tested the approach for short- and medium-range weather prediction, and the work of \cite{Arcomano2023}, which investigated its potential for climate modeling. The hybrid model used for the forecast experiments of the paper is based on the low-resolution, simplified parameterization atmospheric general circulation model SPEEDY. In addition to the hybridized prognostic variables of SPEEDY, the model has three purely ML-based prognostic variables: the 6~h cumulative precipitation, the sea surface temperature, and the heat content of the top 300 m deep layer of the ocean \citep[a new addition compared to the model used in][]{Arcomano2023}. The model has skill in predicting the El Ni\~no cycle and its global teleconnections with precipitation for 3-7 months depending on the season. The model captures equatorial variability of the precipitation associated with Kelvin and Rossby waves and MJO. Predictions of the precipitation in the equatorial region have skill for 15 days in the East Pacific and 11.5 days in the West Pacific. Though the model has low spatial resolution, for these tasks it has prediction skill comparable to what has been published for high-resolution, purely physics-based, conventional, operational forecast models.
\end{abstract}

\section*{Plain Language Summary}
We assess the potential of a hybrid modeling approach for weather prediction beyond 7-10 days. This approach combines machine learning (ML) with a conventional physics-based model of the atmospheric general circulation. Some ML components of the hybrid model make corrections to the variables modeled by the conventional model. Other components provide information about the changes in the sea surface temperature that result from interactions between the atmosphere and ocean, or provide enhanced capabilities for the prediction of the precipitation. The results of forecast experiments with the model show that it has skill in predicting the El Ni\~no cycle and its influence on the weather in other parts of the world for 3-7 months depending on the season. The model also provides useful forecast information about the precipitation in the Tropics for 11-15 days depending on the location. The tested approach could be used both to hybridize current high-resolution purely physics-based operational models and to enhance the capabilities of low-resolution models of the general circulation used in academic research. 

\section{Introduction}

Steady progress in weather prediction beyond the medium (7-10 days) forecast range has been made in numerical weather prediction (NWP) \citep[e.g.,][]{NAP21873,White2022}. Advances in land and ocean modeling have been crucial for these improvements \citep[][]{Brassington2015}. Both the European Centre for Medium-Range Weather Forecasts (ECMWF) \citep[][]{Vitart2014} and the National Centers for Environmental Prediction (NCEP) \citep[][]{CFSv2} have coupled ocean-atmosphere ensemble-based systems, with predictions extending well beyond 10 days lead time. However, the longer range forecasts prepared with these systems tend to have substantial biases \citep[e.g.,][]{zhangs2s,White2017,molod,Mouatadid,Monhart,Domeisen}. Improved modeling of the atmosphere-land-sea-ice interactions and tropical convection (e.g., convection-coupled equatorial waves) has long been considered pivotal for gaining forecast skill beyond the ten days \citep[e.g.,][]{Zhang2013}. Using machine learning-based (ML-based) model components is one promising avenue to achieve this goal, with the additional benefit that such model components can also substantially reduce the computational cost associated with fully coupled, purely physics-based models. 

Current ML-based weather forecasting models have demonstrated skill comparable to, or even surpassing, that of the operational Integrated Forecasting System (IFS) of ECMWF in the 1 day to 10 days forecast range for many atmospheric state variables \citep[e.g.,][]{Bi2023,Lam2023,chen2023fengwu,chen2023fuxi,nguyen2023scaling}. While it is less clear that these ML-based models are also suitable for prediction beyond the medium range \citep[][]{chattopadhyay2023longterm}, there have been some promising initial attempts to show that this may be the case. For instance, \cite{weyn} showed that an ensemble prediction system based on an ML model of 6 prognostic variables provided ensemble mean forecasts that were only modestly inferior to the ECMWF ensemble mean forecasts in the 4 to 6 weeks forecast range. More recently, the FuXi-S2S model, the first ML-only based model designed specifically for forecasting beyond the medium range \citep[][]{chen2023fuxis2s}, was shown to perform on par, or slightly better, than the ensemble mean of the ECMWF system for precipitation anomaly and Madden-Julian oscillation (MJO) forecasts out to 6 weeks. However, this 6 week time scale is not long enough to take advantage of the potential predictability associated with teleconnections of the El Ni\~no. 

The incorporation of machine learning into existing conventional physics-based forecast models is appealing because it takes advantage of both the vast physics-based knowledge built into those models and the availability of decades-long time series of (observation based) reanalysis of the atmospheric states, which can be used for ML training. Hybrid models that implement such an approach have been shown to be stable during decades-long simulations, reduce biases of the host atmospheric general circulation model (AGCM), and improve weather forecasts \citep[][]{Watt-Meyer2021,Arcomano_2022,Arcomano2023,kochkov2023neural,Kwa2023}. \cite{Arcomano2023} (AEA23 hereafter) described a hybrid model in which the atmospheric component was a low-resolution AGCM hybridized with ML, while the oceanic component was a purely ML-based model of the sea surface temperature (SST). This model stably produced realistic atmospheric states and SST fields in a 70-year simulation, significantly reducing the atmospheric biases of the host AGCM, while also better capturing the temporal variability of the atmospheric states. It also reproduced many important features of the observed SST climatology with good accuracy, including the variability associated with El Ni\~no. The variability and low biases of the hybrid model were comparable with those of high-resolution physics-based climate models, though the hybrid model required only modest computational resources to train and run.

Here we present results obtained with a new version of the model of AEA23. This new version of the model has better forecast skill than the original model at forecast lead times longer than 7 days. The new features of the model that lead to this improvement of the forecast skill are an ML-based prognostic variable for the heat content of the top 300 m layer of the ocean, and capabilities to handle disparate time scales of the atmospheric and oceanic dynamics. The decision to include the heat content of the top 300 m layer of the ocean was motivated by physical reasoning \citep[e.g.,][]{McPhaden, Ji, Clarke}. As in AEA23, the AGCM is the Simplified Parameterization, primitive-Equation Dynamics (SPEEDY) model \citep[][]{Molteni,Kucharski}. The low resolution of this AGCM has allowed us to relatively rapidly test many different configurations of the hybrid model.

The structure of the paper is as follows. Section 2 provides a brief summary of the hybrid methodology to help explain the newly added features of the model. A more detailed description of the model and the training procedure are provided in Appendix A. Section 3 discusses the data used in the numerical experiments for both training and assessing the forecast performance of the model. Section 4 demonstrates the forecast capabilities of the model, examining its predictive skill for various atmospheric and oceanic phenomena based on $195$ experimental two years long forecasts. The main result is that the model has useful forecast skill well beyond the medium range. Specifically, the model has skill in predicting El Ni\~no and its global precipitation teleconnections for 3-7 months depending on the season. The model can also maintain variability of the precipitation associated with equatorial Kelvin and Rossby waves and also in the wavelength-frequency range of MJO for the entire two year duration of the forecasts. In addition, the forecasts of the precipitation anomalies in the equatorial region have skill for 15 days in the East Pacific and 11.5 days in the West Pacific. Section 5 presents further discussion and the conclusions.

\begin{figure}
\includegraphics[scale=0.5]{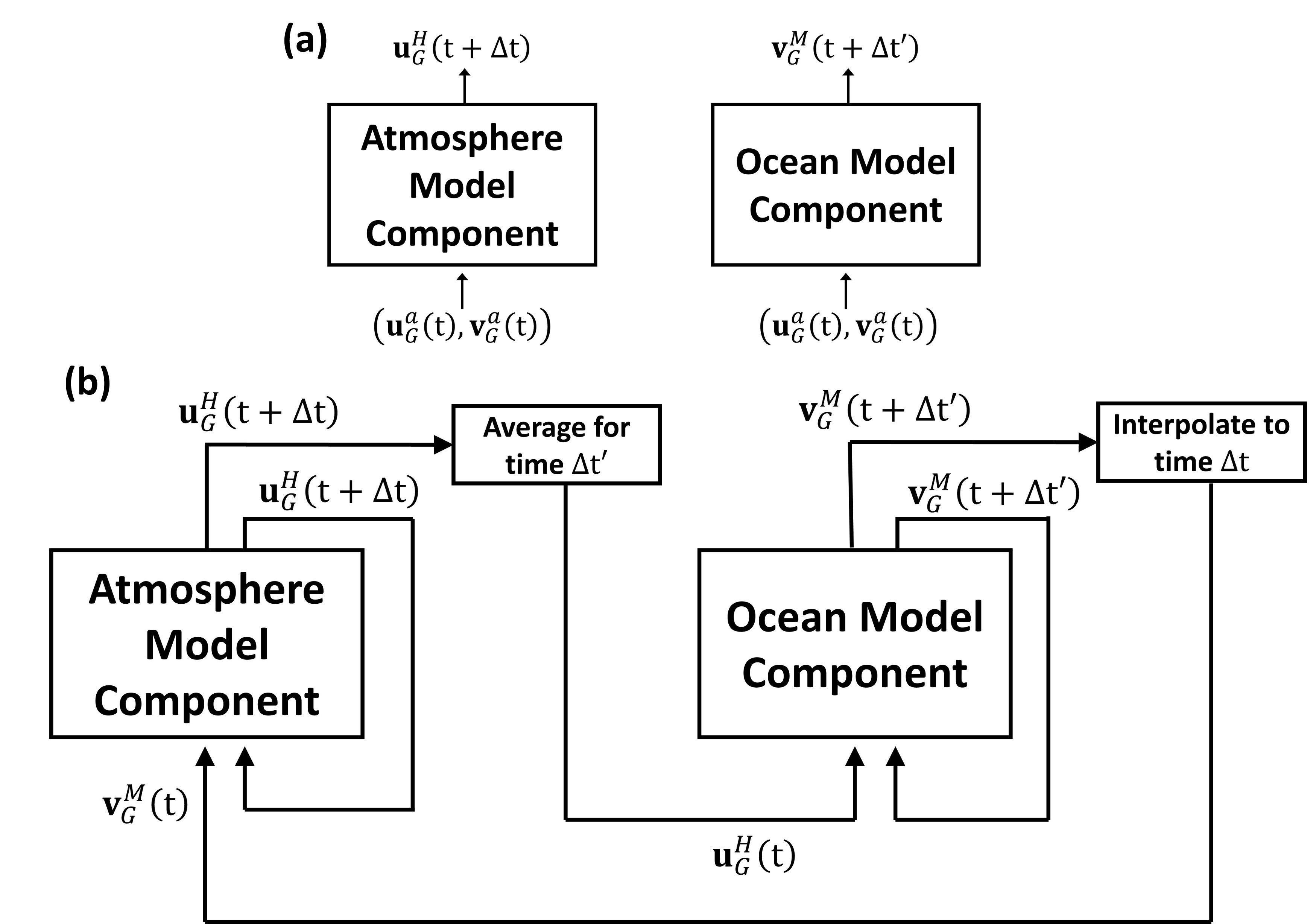}
\caption{Schematic of the model setup during (a) training and (b) prediction. The atmospheric model component and ocean model component are trained independently as shown in (a), but coupled during prediction as shown in (b). We take the atmospheric (oceanic) model state $\mathbf{u}_G(t)$ ($\mathbf{v}_G(t)$) to be $\mathbf{u}_G^a(t)$ ($\mathbf{v}_G^a(t)$) during training and $\mathbf{u}_G^H(t)$ ($\mathbf{v}_G^M(t)$) during prediction. The super-script $a$ denotes the observation-based analysis atmospheric or oceanic states, and the super-script $H$ ($M$) denotes the states predicted by the hybrid atmospheric (ML-only oceanic) model component. Since the time step $\Delta t$ of the atmospheric model component is taken to be smaller than the time step $\Delta t'$ of the oceanic model component (i.e.,
$\Delta t'{=}n\Delta t$), the atmospheric model component variables used as input to the oceanic model component are time-averaged over a trailing $n\Delta t$ time window. In a similar fashion, the oceanic model component variables fed to the atmospheric model component are interpolated in time to produce the $\Delta t$ time step of the oceanic input to the atmospheric model component. See Appendix A for more details.}
\label{fig:arch_train_pred}
\end{figure}

\section{Methods}\label{sec:Methods}
Our model has the same hybrid architecture, and it is also trained by the same protocol, as the model of AEA23. It has two major components: (1) a hybrid component for the atmosphere that combines SPEEDY with an ML model, and (2) an ML-only component for the ocean. We denote the vector of the global atmospheric prognostic state variables by ${\bf u}_G(t)$ and the vector of the global oceanic prognostic state variables by ${\bf v}_G(t)$. The two model components are coupled by using both ${\bf u}_G(t)$ and ${\bf v}_G(t)$ as input to both of them (see Fig. \ref{fig:arch_train_pred}). Though they are trained separately to make one-time-step forward predictions ${\bf u}_G(t+\Delta t)$ and ${\bf v}_G(t+\Delta t')$, they interact nonlinearly in the predictions, in which the outputs ${\bf u}_G(t+\Delta t)$ and ${\bf v}_G(t+\Delta t')$ serve as the inputs of the next time step (Fig. \ref{fig:arch_train_pred}). Note that the time steps $\Delta t$ of the atmospheric model component and $\Delta t'$ of the oceanic model component may be different. For the results we report, we use time step $\Delta t = 6$ hours for the hybrid atmospheric component and $\Delta t' = 72$ hours for the ML-only oceanic component (both are much longer than the finite differencing time step of SPEEDY). The details of the model architecture and its training are discussed in Sec. \ref{sec: ML Model  Dynamics} of the Appendix. 

The model grid of both model components is divided into local sub-domains, and the state variables at every grid point within a local sub-domain are modeled by an individual and independently trained reservoir computer (Jaeger, 2001; Lukosevicius and Jaeger, 2009; Lukosevicius, 2012). This architecture allows the ML model training to be parallelized (Pathak, Hunt, et al. 2018), therefore effectively utilizing the resources of a high performance computing environment. For the atmospheric component, each sub-domain has a rectangular base of 4 latitude-longitude grid points and includes all vertical levels. The vertical grid consists of eight $\sigma$-levels (defined at $\sigma = 0.025$, $0.095$, $0.20$, $0.34$, $0.51$, $0.685$, $0.835,$ and $0.95$), where $\sigma$ is the ratio of pressure to the surface pressure. For the ocean model, each sub-domain is just the rectangular base of an atmospheric subdomain (i.e., the ocean model is two-dimensional). The blue box in Fig. 2 denotes a rectangular base of a sub-domain. In order to model the state variables at grid points in its sub-domain, a reservoir computer receives as input the information from grid points in the immediate neighborhood of its sub-domain (shown by the red box in Fig. 2). The atmospheric hybrid component is constructed according to the Combined Hybrid-Parallel Prediction (CHyPP) approach for modeling large and complex spatiotemporally chaotic systems \citep[][]{Pathak+Wikner,Wikner2020}, in which a single global knowledge-based model (in our case, SPEEDY) is combined with a data-driven ML model composed of locally-connected individual reservoir computers. See \ref{sec: ML details} for a more detailed discussion of the model architecture.

\begin{figure}
\begin{center}
\includegraphics[scale=0.5]{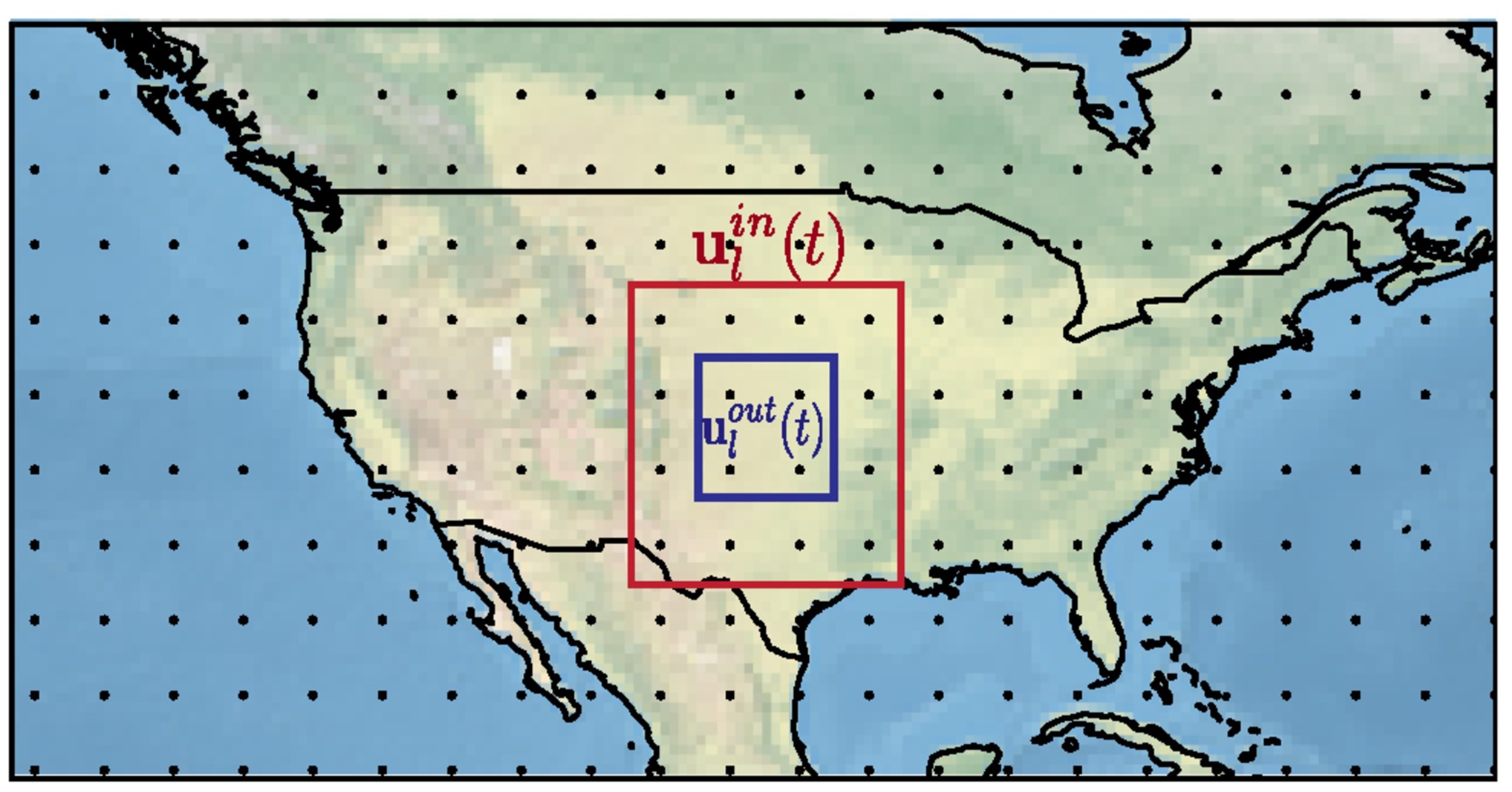}
\caption{Illustration of the geographic domain decomposition for a local input vector $\mathbf{u}^{(in)}_l$ (with grid points enclosed by a red box), and a local output vector $\mathbf{u}^{(out)}_l$ (with grid points enclosed by a blue box). The black dots correspond to grid points of the SPEEDY model.}
\label{fig:domain_decomp}
\end{center}
\end{figure}

\subsection{Design of the Ocean Component}\label{sec:Exp Design}
The ML component of the atmospheric hybrid model as well as the ML-only ocean model consist of two types of parameters: a set of learnable parameters that are determined through the training procedure, and a set of hyperparameters that are chosen a priori and refined by experimentation (model ``tuning"). Here we will discuss some of the experimentation that went into choosing the hyperparameters of our model components (refer to Sec. \ref{sec: Model Training} for detailed information regarding the training procedure of our model). We will focus more on the design choices for the ML-only ocean component, which introduces several modifications over that of AEA23. For the hybrid atmospheric component, we keep the same hyperparameters as those used in AEA23. For the oceanic model component, we use the same hyperparameters for reservoir computers across all sub-domains. We found that varying some of the hyperparameters with latitude like for the atmosphere does not improve the ability of the oceanic model component to predict the SST. Each sub-domain is modeled with a reservoir computer of $N'=6000$ nodes. We found that using a reservoir with a larger $N'$ did not significantly improve the forecast skill of the model. Furthermore, we found that the skill of the oceanic model component was generally not very sensitive to the choice of many hyperparameters such as the spectral radius $\rho'$ and the degree of the reservoir adjacency matrix, $d'$. On the other hand, we found that the forecast skill of the model was more sensitive to the choice of hyperparameters such as the $L2$ regularization strength $\beta'$ (also called the ``Tikhonov regularization") of the model's training loss function and leaking rate parameters $c'_i$ for $i=1,...,N'$ (which control the amount of memory retained in the next reservoir node state of the previous node states). The Tikhonov regularization strength was tested in powers of 10 from $10^{-4}$ to $10^3$, and we found that for values lower than $1$, the model was significantly more likely to become unstable during prediction. We selected the leaking rate for each reservoir node by randomly choosing values from a log-uniform distribution over the interval $[10^{-3},1]$ like \cite{Tanaka}. We found that using multiple leaking rate values (i.e., as opposed to choosing a single leaking rate $c'_i=c$) was crucial to the ocean model to simultaneously predict both the SST and the 300~m oceanic heat content fields with realistic multi-year variability. (See Table \ref{tab:ocean_HP} for the list of chosen hyperparameter values.)

The inclusion of 300 m ocean heat content as a prognostic variable in the oceanic model component is motivated by the physical reasoning that the knowledge of the upper-oceanic heat content can improve predictions of the El Ni\~no cycle, especially across the so-called ``spring predictability barrier" \citep[][]{McPhaden, Ji, Clarke}. As expected, this modification noticeably improves the overall forecast performance of the model, including an extension of the skill of the model to predict the El Ni\~no cycle by at least $2$ months. For similar reasons, we also tried adding the ocean mixed layer depth as a prognostic variable to the model, but found that it did not improve the performance of the model. We did tests (only cursory, however, due to computational constraints) of the effects of including only some atmospheric variables of the lowest atmospheric model level in the input to the ocean model. We added, one-by-one, the temperature, the two components of the horizontal wind vector, and the specific humidity of the lowest atmospheric model level, and (the natural logarithm of) the surface pressure. We found that the addition of each variable contributed to obtaining realistic temporal variability and/or extending the forecast horizon of the model-predicted SSTs. In addition, we tried inputting a signal that consisted of the sine and cosine of the day of the year multiplied by $2\pi / 365$ to the ocean (and atmospheric) model in order to encode knowledge of the time of the year. This did not, however, appear to provide any performance benefits. The time step $\Delta t'$ for the ocean model is also a hyperparameter and we tested time steps of $1$-day, $3$-days, $7$-days, and $1$-month. We found that the $3$-day time step resulted in the best overall model performance. While the $7$-days and $1$-month time steps also produced good results, the $1$-day time step resulted in noticeably degraded performance (possibly due to the choice of other hyperparameters). Some of our other tests included having an input-pass-through (i.e., the input to the reservoir is also provided as an input directly to the output layer), training to predict tendencies rather than the state itself (i.e., $\delta \mathbf{x}(t+\Delta t) = \mathbf{x}(t+\Delta t) - \mathbf{x}(t)$ instead of $\mathbf{x}(t+\Delta t)$), and separate Tikhonov regularization strengths for the input-pass-through component and the reservoir state component of the output feature vector. None of these methods appeared to improve the model performance (and in some cases they degraded it), and were therefore left out of the final version of our model.

\section{Data}\label{sec:Data}
The global analyses for the atmospheric variables and SST are hourly ERA5 reanalyses \citep[][]{Hersbach} interpolated spatially to the SPEEDY computational grid using a $2$-dimensional quadratic B-spline interpolation in the horizontal direction and a 1-dimensional cubic B-spline interpolation in the vertical direction (to fit to the eight prescribed constant values of $\sigma$). The global analyses for the 300 m oceanic heat content are the monthly mean ORAS5 reanalyses  \citep[][]{ORAS5} interpolated spatially to the SPEEDY computation grid as described for the ERA5 reanalyses. In addition, the ORAS5 data is linearly interpolated to hourly data in the temporal domain. The training starts at $0000$ UTC on $1$ January $1981$ and ends at $0000$ UTC on $27$ December $2002$ ($K\approx 1.93\times 10^5$).

\section{Results}\label{sec:Results}
The results presented in this section for the assessment of the forecast skill of the hybrid model are based on $195$ two-year long forecasts, with the initial forecast times equally spaced by $30$ days between $0000$ UTC $16$ January $2003$ and $0000$ UTC $23$ December $2018$. We evaluate the performance of the model by verifying the forecasts against ERA5 reanalyses. Where appropriate, we also present performance metrics for persistence and climatology based forecasts for comparison. By persistence based forecast, we mean using the analysis that serves as the initial condition of the hybrid model forecast to provide the forecast for all lead times, that is, keeping the forecast state constant in time. The climatology based forecast uses the climatological values of the prognostic state variables for the specific location and calendar time of the year. We also present verification metrics for members of the North American Multi-Model Ensemble (NMME), which are state-of-the-art physics-based model forecasts. We obtained these metrics by downloading the publicly available NMME forecast data \citep[][]{Kirtman} and using the same computer codes for the computation of the verification metrics as for the hybrid model.

We calculate verification metrics for forecast parameters that are all scalar-valued functions of the prognostic state variables. The two metrics we use are the root-mean-square error (RMSE) of the forecast parameters and the Pearson correlation coefficient (PCC) between the values of the forecast parameter for the forecasts and the ERA5 reanalyses. (See \ref{subsec:metrics} for the formal definition of the two metrics). When we show the PCC in a map, we calculate its values at each spatial grid point independently. We also compare the climate associated with the hybrid model forecasts to that represented by the ERA5 reanalyses.  In these comparisons, we compare a total of 390 years of hybrid forecasts with about 18 years of ERA5 data (the difference in sample sizes creates some minor artifacts).

\subsection{El Ni\~no Cycle and its Teleconnections}
The El Ni\~no cycle is the dominant driver of year-to-year global climate variability \citep[e.g.,][]{McPhaden1,Lin,Timmermann}. It is a coupled ocean-atmosphere phenomenon \citep{Bjerknes} with three phases, which are characterized by the presence of anomalously warm (El Ni\~no), anomalously cold (La Ni\~na), or normal (neutral) sea surface temperatures in the central and eastern tropical Pacific Ocean. El Ni\~no or La Ni\~na conditions are usually accompanied by anomalous atmospheric weather patterns across the globe, such as increased or decreased amounts of rainfall \citep[e.g.,][]{Lin,Wang,Losada,Dai}. Thus predicting the El Ni\~no cycle holds significant socioeconomic and environmental value \citep[e.g.,][]{Liu}. Next, we describe the skill of our model in predicting this phenomenon and some of its teleconnections around the globe.

\subsubsection{Ni\~no $3.4$ Index}
\begin{figure}
\includegraphics[scale=0.8]{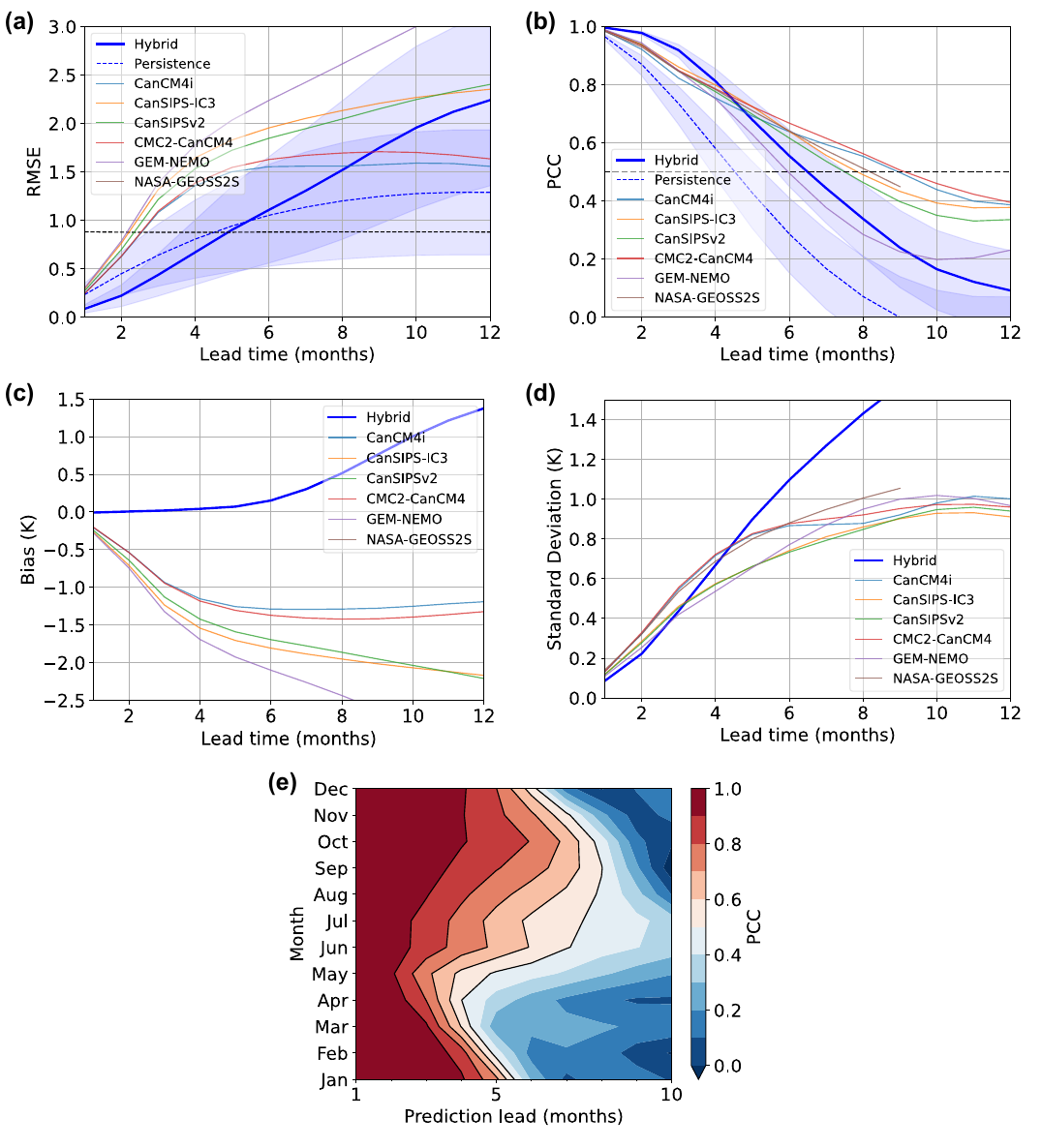}
\caption{Illustration of the dependence of the skill of the hybrid model in predicting the Ni\~no 3.4 Index on the forecast lead time. Shown are the (a) RMSE (b) PCC (c) forecast bias, and (d) standard deviation of forecast errors for (thick solid blue) the hybrid model and (thin solid various colors) various state-of-the-art conventional physics-based models. For reference, black dashes show the RMSE for the climate based forecasts in (a) and the PCC=$0.5$ line in (b). Also shown is (e) the dependence of the PCC for the hybrid model on the (x-axis) forecast lead time and (y-axis) month of the start of the forecast.}
\label{fig:enso_pcc_rmse}
\end{figure}

The phase of the El Ni\~no cycle is usually characterized by the Ni\~no $3.4$ Index defined by the spatially averaged monthly mean SST anomalies in the Ni\~no $3.4$ region ($5\degree$N-$5\degree$S, $170\degree$W-$120\degree$W). Figure~\ref{fig:enso_pcc_rmse} shows the forecast skill of our model in predicting this index. The top two panels present the RMSE and PCC metrics as functions of forecast lead time for the different forecasts. In terms of the RMSE (panel a), the hybrid model (thick solid blue line) performs better than the state-of-the art physics-based models (thin solid lines of various colors) up to about 8.5\,months.
In addition, the RMSE for the hybrid model forecasts reaches the value for the climatology-based forecasts (horizontal dashed black line) at about 5 months lead time, and the value for the persistence-based forecast (thin dashed blue line) at about 5.5~months. In contrast, the RMSE for the physics-based NMME models reaches the value for the climatology-based forecasts at about 2-2.5~months and remains higher than that for the persistence-based forecasts at all lead times. The PCC (panel b) for the hybrid model is higher than those for the NMME models up to 4\,months and it is also higher than that for the persistence-based forecasts at all lead times. The PCC drops to $0.5$ (horizontal black dashed line) for the hybrid model at about 6.5\,months and for the NMME models between about 6\,months and 9.5\,months. In summary, the performance of the hybrid model is superior to those of the NMME models with respect to both metrics up to 4\,months, and with respect to the RMSE up to 8.5\,months. Beyond that time, the NMME models perform better than the hybrid model with respect to both metrics.

To further investigate the change in the relative performance of the hybrid model with the forecast lead time, we decompose the RMSE into a bias and a standard deviation component (the sum of the square of the bias and the square of the standard deviation is equal to the square of the RMSE). The results show (Fig.~\ref{fig:enso_pcc_rmse}, panels c and d) that the shorter term superiority of the hybrid model is the result of two factors: first, the bias (panel c) is practically negligible for the model up to about 5\,months, while it is rapidly growing for the NMME models in the same forecast range; second, the standard deviation of the errors for the model does not become larger than those for the NMME models until about 3.5-4\,months. However, both the bias and the standard deviation of the errors keep growing beyond 5\,months for the hybrid model, while they tend to saturate for the NMME models. The growing standard deviation of the errors for the hybrid model is the result of the model having difficulty in maintaining the neutral phase of the El Ni\~no cycles.
This result suggests that while the version of the model described in the present paper has more skill in predicting the weather beyond the medium forecast range than that of AEA23, it is less appropriate for long-term climate simulations, because the AEA23 version of the model produced more realistic El Ni\~no cycles in a 70 year long climate simulation. 

The last panel of Fig.~\ref{fig:enso_pcc_rmse} (panel e) illustrates the seasonal dependence of the skill of the hybrid model in predicting the El Ni\~no cycle. In this panel, the x-axis is the forecast lead time, while the y-axis indicates the month in which a forecast is started. If the predictability horizon is defined by the lead time at which the value of the PCC drops to PCC=0.5, the model can predict the El Ni\~no cycle with a lead time from 4\,months in the spring to 8\,months in late summer and early fall. The low predictability of the El Ni\~no cycle in the spring is a manifestation of the ``spring predictability barrier" \citep[e.g.,][]{Zheng,Lai}. 

The superior performance of the hybrid model to that of the state-of-the art physics-based models for the first 4 forecast months is achieved while requiring substantially more modest computing resources to prepare the forecasts. Once trained, the bulk of the computing cost of our hybrid model can be attributed to frequently restarting the physics-based component (SPEEDY), which is a low-resolution global atmospheric model. Evidently, despite using a low-resolution model as the physics-based component of our hybrid model, which is computationally cheaper to operate than the high-resolution coupled models like those used in NMME, we are able to obtain results comparable to computationally more expensive physics-based models. 

A second class of forecast models we can compare our hybrid model to are purely data-driven models \citep[e.g.,][]{Ham,Zhou}. While such models have more skill in predicting the Ni\~no $3.4$ Index, they are highly-specialized models that are either directly trained to predict the index, or only provide spatiotemporal forecasts in a small geographical region (e.g., the equatorial Pacific Ocean). Furthermore, while such data-driven models typically have a temporal resolution of one month, the ocean component of our model is an auto-regressive global forecasting model with a temporal resolution of $72$ hours. Despite having a temporal resolution much finer than the time scale over which El Ni\~no evolves (on the order of months), our model has useful El Ni\~no forecasting skill. Several machine learning models that predict the global atmospheric and oceanic states similar to our model have recently been developed \citep[e.g.][]{ola,chen2023fuxis2s}. However, a direct comparison with these models is not feasible either because the long-term predictions needed to evaluate  El Ni\~no forecast have not been published \citep{chen2023fuxis2s}, or the common metrics of RMSE and PCC are not reported \citep{ola}.

\subsubsection{Teleconnections}
ENSO influences global circulation and weather patterns beyond the medium forecast range via teleconnections \citep[e.g.,][]{Alexander,Lin}. For instance, El Ni\~no (La Ni\~na) is typically accompanied by drier (wetter) conditions in the eastern Pacific and over the maritime continent, while ushering wetter (drier) conditions over the southern United States \citep[e.g.,][]{Ropelewski,Mason}. The atmospheric response during El Ni\~no events also influences, via the ``atmospheric bridge", ocean basins across the globe \citep{Alexander}. To assess whether our model is able to capture these global teleconnections at the seasonal time scale, we investigate the ability of our model to capture the observed seasonal correlations between El Ni\~no and other atmosphere-ocean state variables (e.g., SSTs and precipitation) across the globe. 

\begin{figure}
\includegraphics[scale=0.6]{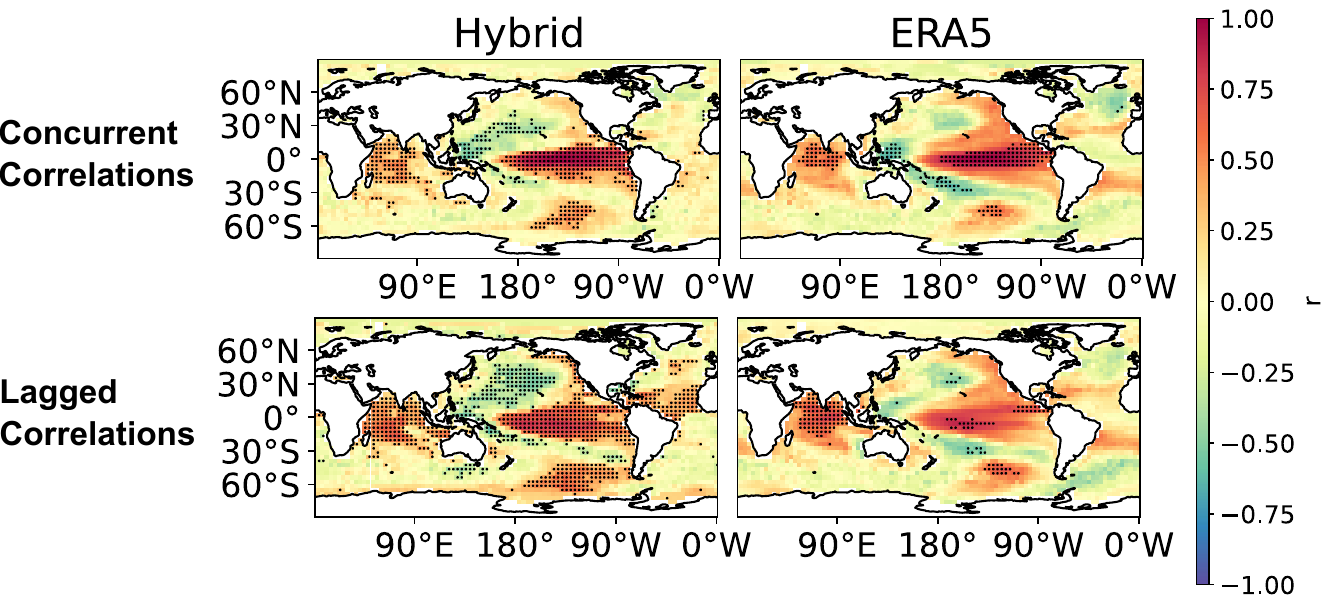}
\caption{The skill of the seasonal hybrid model forecasts in capturing the teleconnection between the Ni\~no 3.4 Index and the global SST anomalies. Shown are the (top) concurrent correlations and (bottom) lagged correlations with a 3-month lag for the SST anomalies for the (left) hybrid model and (right) ERA5 reanalyses. Black dots indicate locations where the PCC is statistically significant at the $95\%$ confidence level.}
\label{fig:enso_sst_corr}
\end{figure}

Figure \ref{fig:enso_sst_corr} shows the correlations between the 3-month averaged SST anomalies at each grid point and the Ni\~no 3.4 Index. The top row of the figure shows concurrent correlations between the SST anomalies and the Ni\~no 3.4 Index, while the bottom row shows lagged correlations between the two quantities. For the calculation of the lagged correlations, we use the Ni\~no 3.4 Index for three-month periods December-January-February (DJF), March-April-May (MAM), June-July-August (JJA), or September-October-November (SON) and the SST anomalies for the following three months. Panels in the left and right column of the figure show the correlations for the hybrid model and the ERA5 reanalyses, respectively. The results show that the hybrid model predictions capture the observed strong positive correlations between the Ni\~no 3.4 Index and SST stretching from the central and eastern equatorial Pacific Ocean up (down) to the western coast of North (South) America. In addition, the hybrid model captures the ``horseshoe" shaped negative correlations in the equatorial western Pacific, northwestern Pacific and southwestern Pacific. The hybrid model also captures the strong positive correlations in the Indian Ocean. We note, however, that in the lagged correlations, the hybrid model somewhat overemphasizes the positive correlations in the central Atlantic and the negative correlations in the northwestern Pacific. It also slightly under-emphasizes the spatial extent of the positive correlations off the western coast of North America. Aside from these small difference, the hybrid model is able to capture the general large-scale correlations between the SST anomalies in the tropical Pacific and the rest of the world, which is a key component of the El Ni\~no teleconnections \citep[e.g.,][]{Alexander,Lin,Roy}.

\begin{figure}
\includegraphics[scale=0.65]{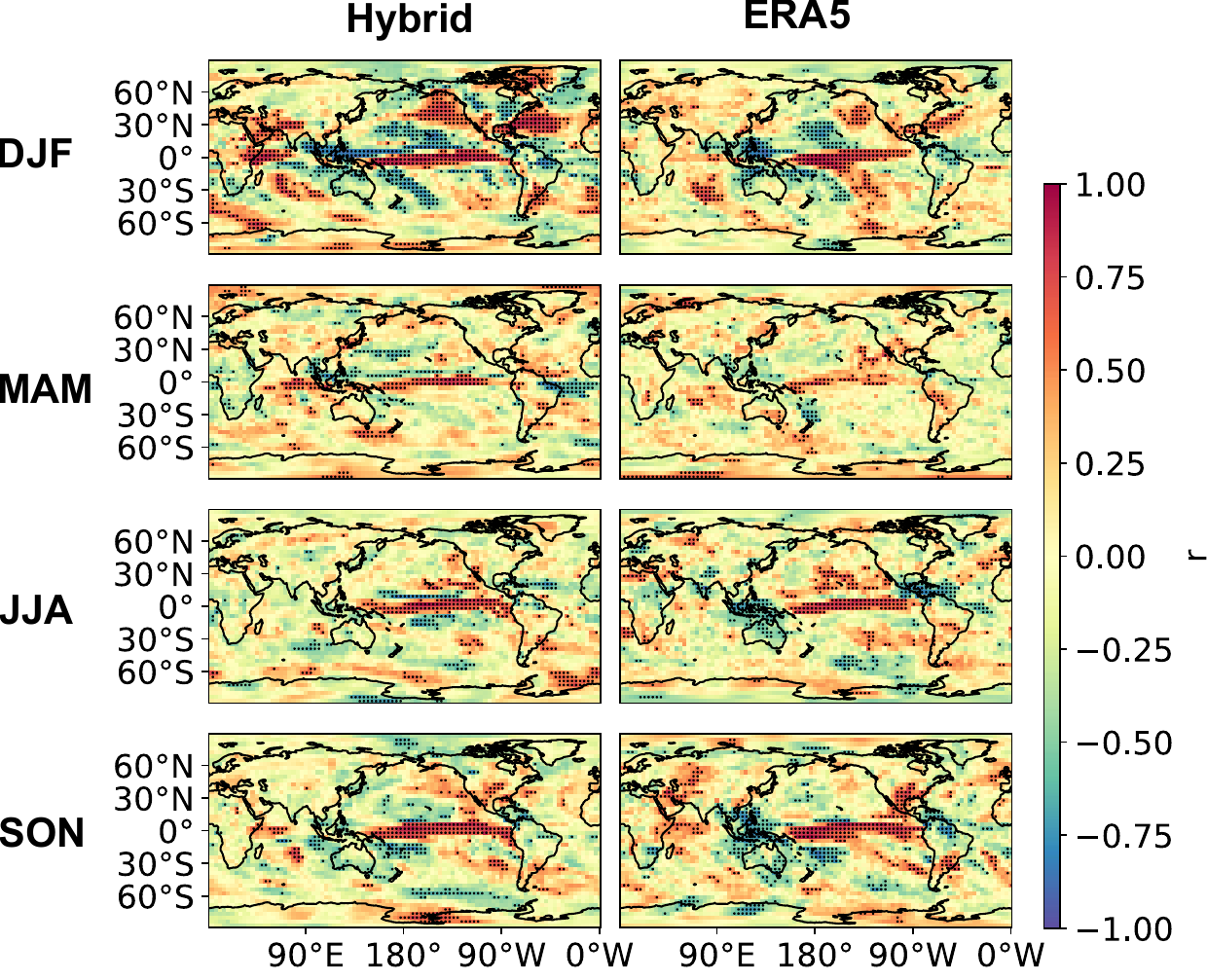}
\caption{The skill of the seasonal hybrid model forecasts in capturing the teleconnections between the El Ni\~no cycle and the precipitation anomalies around the globe. Shown are the concurrent PCC for the Ni\~no 3.4 Index and the seasonal precipitation anomalies for the (left) 3-month lead time hybrid model forecasts and (right) ERA5 reanalyses. The verification times of the forecasts fall in the (from top to bottom) DJF, MAM, JJA, and SON season. Black dots indicate the locations where the PCC is significant at the $95\%$ level.}
\label{fig:enso_precip_corr}
\end{figure}

Figure \ref{fig:enso_precip_corr} shows concurrent correlations between the seasonal precipitation anomalies and the Ni\~no $3.4$ Index. The left panels show the correlations for the one season (3 months) long hybrid model forecasts, while the right panels show the corresponding fields for the ERA5 reanalyses. The different rows show results for different seasons. For the hybrid model, the seasons correspond to the verification times of the forecasts. For instance, the results shown for DJF are based on the forecasts started in SON. While our forecasts are for $2$ years in length, for the preparation of this figure, we only use the forecasts from the first three months. We placed a black dot on each grid point for which the calculated correlation value is statistically significant at the $95\%$ confidence level calculated using the Fisher transformation \citep{fisherz}. El Ni\~no activity peaks in the DJF period and the resulting strong El Ni\~no-precipitation correlations for the ERA5 reanalyses are shown in the top right panel. As seen in the corresponding top left panel for the hybrid model, the forecasts capture thes strong positive correlations for the precipitation anomalies in the central/eastern Pacific Ocean, northeastern Pacific Ocean, western Atlantic Ocean along the United States, and over the central Indian Ocean. Likewise our model captures the strong negative correlations between the tropical Pacific SST anomalies and precipitation anomalies over the maritime continent, the Pacific Ocean to the north and south of the Ni\~no 3.4 region, and over the northeastern South America. While our model over-emphasizes the strength of some of these correlations (e.g., the positive correlation over the western Atlantic Ocean along the eastern United States), it generally captures the correct sign of the correlation and also has realistic spatial patterns of the correlations. However, our model struggles in some regions. For instance, it incorrectly predicts strong positive correlations over the Arabian peninsula and northern Indian sub-continent during the DJF period. It also struggles to capture the strong negative correlations for the summer monsoon activity over the Indian sub-continent and the maritime continent. Overall, our model forecasts generally contain El Ni\~no-precipitation correlation patterns that resemble those observed in the ERA5 data set, particularly over regions of significantly strong positive and negative correlations. 

\subsection{Global Precipitation}
An important task for models that forecast weather beyond the medium range is to correctly predict precipitation anomalies at lead times ranging from several days to several weeks. Next, we investigate whether our model has useful skill for this task. We do so by applying the PCC metric to weekly precipitation anomalies, as well as by examining directly  the average errors in the weekly-accumulated precipitation forecasts. We note that PCC is a commonly used deterministic verification metric for this task \citep[e.g.,][]{Andrade,Li,ZhaoT}. It is also important to emphasize that even though each ERA5 analysis is based on millions of observations, these observations do not include precipitation observations for most of the globe. The only exceptions are composite ground-based radar/rain-gauge precipitation estimates over the United States east of the Rockies from 2009 onward \citep{Hersbach}. Thus the ERA5 precipitation analyses primarily reflect the precipitation response of the ECMWF Integrated Forecasting System (IFS) Cy41r2 conventional model, which was used in ERA5 to obtain the background fields, to the assimilation of the observations of other state variables. The ERA5 precipitation analyses have the largest errors in the tropics \citep[e.g.,][]{Lavers}. But, because the precipitation prognostic variable of our model was trained on ERA5 precipitation data, the precipitation diagnostics presented here are good measures of the skill of the proposed modeling approach to learn from precipitation training data.

\begin{figure}
\includegraphics[scale=0.15]{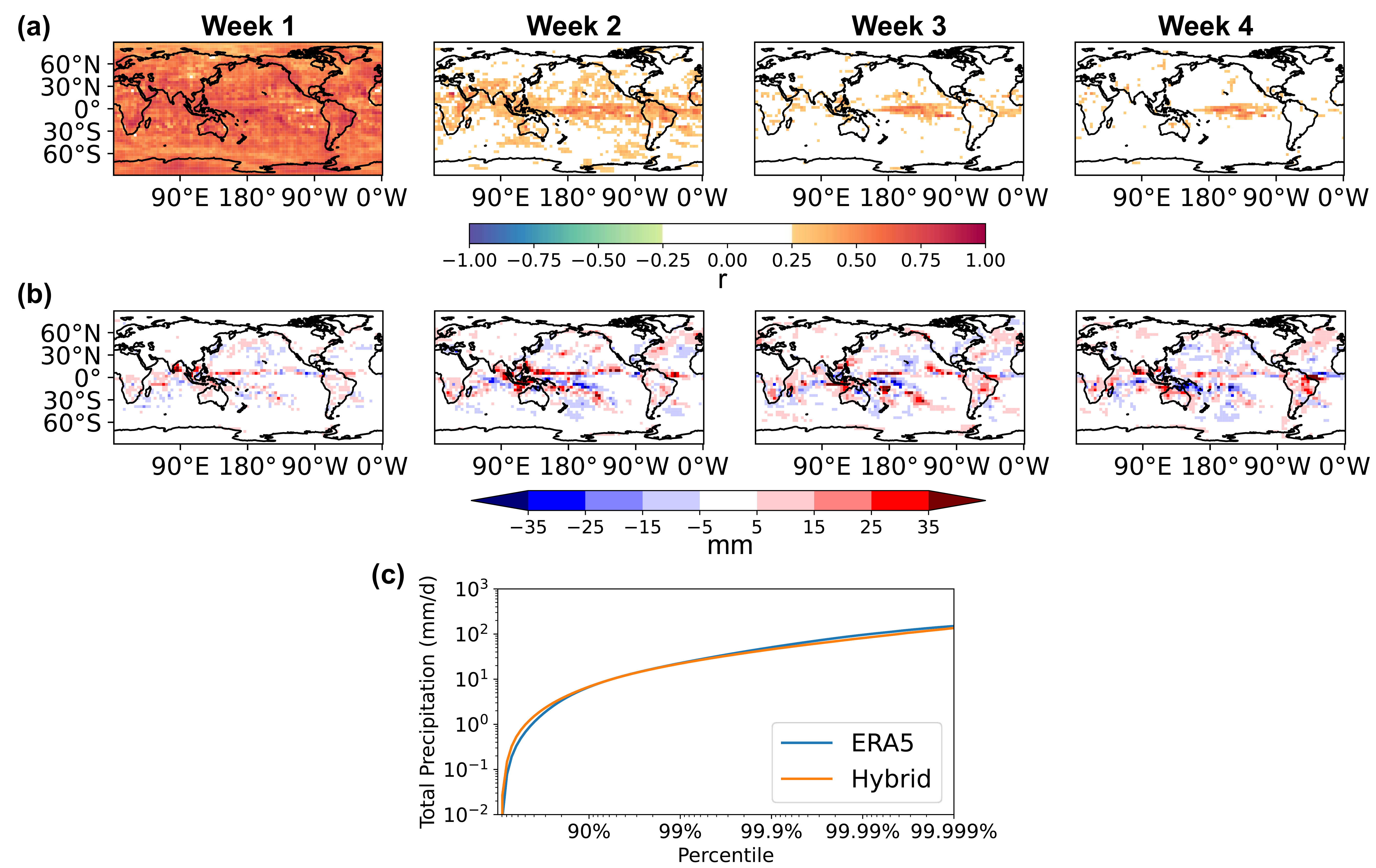}
\caption{The skill of the hybrid model in predicting the anomalies of the weekly mean precipitation.  Shown are the all-season (a) PCC for the weekly mean precipitation between the model forecasts and the ERA5 reanalyses and (b)  difference of the weekly accumulated precipitation between the hybrid model forecasts and ERA5 reanalyses. Results are shown (from left to right) for the week 1, 2, 3, and 4 forecast lead time. Also shown are (c) the rates of occurence of  the different daily precipitation amounts for the (orange) hybrid model forecasts and (blue) ERA5 reanalyses.}
\label{fig:precip}
\end{figure}

Panel (a) in Fig.~\ref{fig:precip} shows spatial correlations between the weekly-mean precipitation anomalies in our model forecasts and the ERA5 reanalyses. There are strong positive correlations across much of the globe at a lead time of $1$ week, but the correlations decay rapidly afterwards. Strong positive correlations persist mostly near the tropics up to a lead time of $4$ weeks. The spatial correlation patterns in our results share similar features to those obtained from some of the models in the S2S prediction project (e.g., Figure $1$ in \cite{Andrade} as well as results in \cite{Li} and \cite{ZhaoT}). There are also regions where our model struggles even at the $1$ week lead time, for instance, over central and western Africa and western South America, but these appear to be similarly problematic regions for many state-of-the-art models, including those in the S2S prediction project \citep{Andrade}. Figure \ref{fig:precip} panel (b) shows the errors in the weekly-accumulated precipitation forecasts at lead times of $1, 2, 3,$ and $4$ weeks, averaged over forecasts made in the DJF period. The overall biases of our model are very small at a lead time of $1$ week, and they remain relatively small even at a lead time of $4$ weeks, especially in the extra-tropics. Figure \ref{fig:precip} also shows that while our model has a slight tendency to both under-predict higher intensity rainfall and over-predict lower intensity rainfall compared to ERA5, it still captures the distribution of rates of rainfall across different percentiles fairly well. Overall, our model appears to have comparable skill to conventional models \citep[e.g.][]{Andrade} in predicting weekly global precipitation anomalies for the first four forecast weeks. 

\begin{figure}
\includegraphics[scale=1]{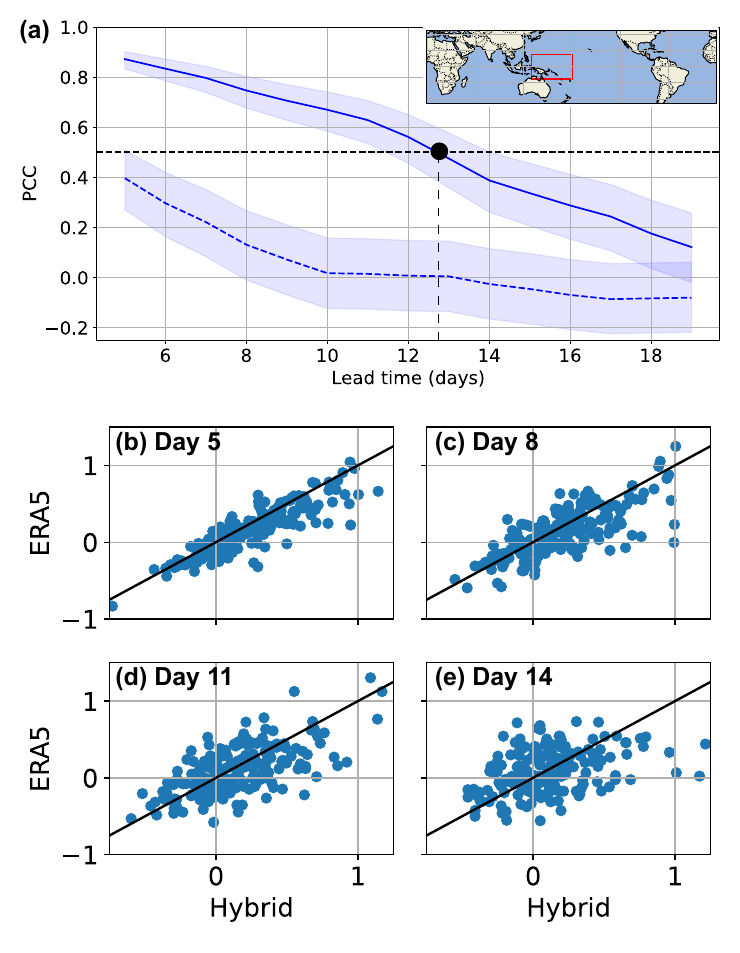}
\caption{Precipitation forecast verification results for the Tropical East Pacific (see red box in the inset of (a)). Shown is (a) the PCC versus forecast lead time for (solid blue) our model forecasts and (dashed blue) the persistence-based forecasts. Blue shading corresponds to a $95\%$ confidence interval of the PCC. The black dashes indicate PCC=0.5. Also shown are scatter plots of the spatially-averaged $5$-day accumulated precipitation anomalies for (horizontal axis) our model predictions and (vertical axis) the ERA5 data at forecast lead times (b) $5$, (c) $8$, (d) $11$, and $14$ days, respectively. The solid black lines have slope $1$, and correspond to a perfect prediction by the hybrid of the observation in ERA5.}
\label{fig:precip_EP}
\end{figure}

\begin{figure}
\includegraphics[scale=1]{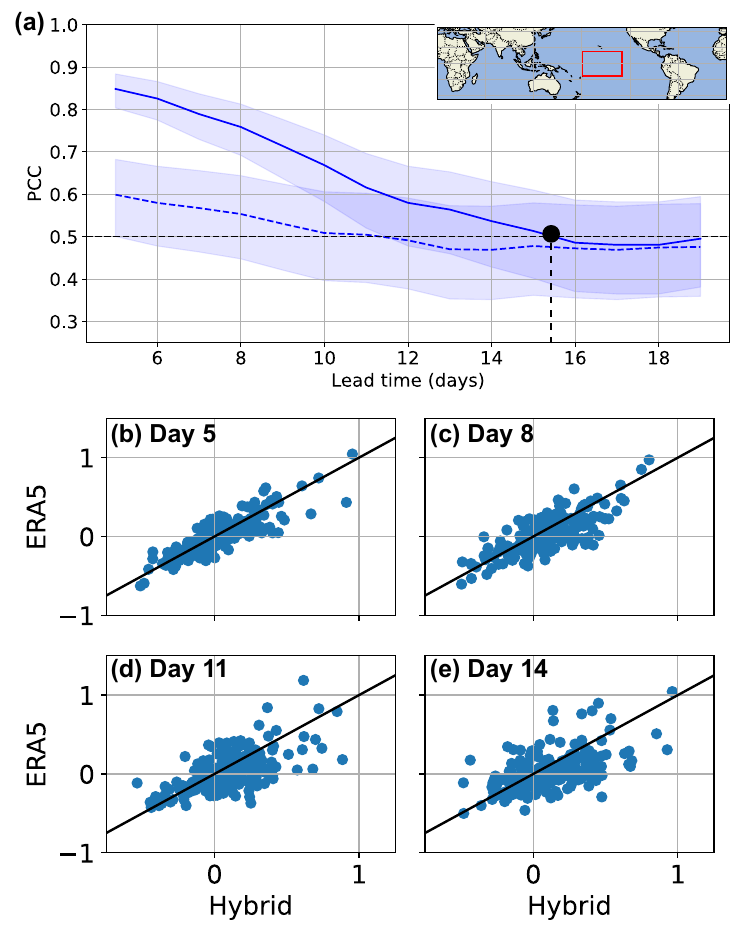}
\caption{Same as Figure~\ref{fig:precip_EP}, except for the Tropical West Pacific (see red box in the inset of (a)).}
\label{fig:precip_WP}
\end{figure}

Rainfall anomalies in the tropical Pacific can impact weather and climate across the globe, making the prediction of tropical rainfall an important task for models. Current models suffer from substantial biases in this region \citep[e.g.][]{lee,sobel,you2024}. To evaluate the skill of our model in predicting rainfall in the tropical Pacific, we calculate verification metrics for two geographical regions: the East Pacific ($15\degree S - 15\degree N$, $180\degree W - 130\degree W$), and the West Pacific ($15\degree S - 15\degree N$, $130\degree E - 180\degree E$). \cite{you2024} suggested separating these two regions and our verification results (Fig. \ref{fig:precip_EP} and Fig. \ref{fig:precip_WP}) also confirm that they pose different forecast challenges. These results are for the prediction of precipitation anomalies, which are calculated as follows.
We first obtain the daily accumulated precipitation anomalies for both our forecasts and ERA5 by subtracting the corresponding ERA5-based day-of-the-year accumulated precipitation climatological mean. Next, for each day in the forecast (and the corresponding verification ERA5 data), we calculate a trailing 5-day sum of the accumulated precipitation anomalies. Finally, we perform a spatial average over the region of interest to obtain a uni-variate time series of the $5$-day accumulated precipitation anomalies. Panel (a) of both figures shows the PCC as a function of lead time for our model forecasts and the forecasts based on persistence. In the East Pacific region, the PCC initially (for the first $5$ days) is higher than 0.5, and as the forecast lead time increases, it converges to about 0.5 for both forecasts. The result that the PCC converges to 0.5 rather than zero indicates that in this region the anomalies typically have a component that changes at time scales that are substantially longer than the longest (19 days) lead time in the figures. Nevertheless, our model predicts the part of the anomalies that affects the forecast accuracy in the investigated range more accurately than persistence for about $15$ days. Predicting the anomalies is more challenging in the West Pacific, where the PCC for the persistence based forecast is already lower than 0.5 at initial time, and it drops to zero by 10 days lead time, showing that the anomalies change fast in this region. Our model does well in predicting these anomalies, greatly outperforming the persistence-based forecasts. The PCC drops to 0.5 at about 12.5 days lead time and it does not reach zero even at 18 days lead time. Panels (b-e) show scatter plots of the predicted (horizontal axis) and ERA5-based (vertical axis) precipitation anomaly values used for the calculation of the PCC. These plots suggest that the anomalies tend to have positive values more often than negative values. This can be explained by a shift in the distribution of spatially-averaged $5$-day accumulated precipitation values in ERA5 that occurs around $2002$ (not shown here). For example, the ERA5 average $5$-day accumulated precipitation (spatially-averaged over the WP region) increased by about $9\%$ for the time period of $2003-2020$ from its average value over the time period of $1981-2002$. We used the ERA5 data corresponding to the period $1981-2002$ (i.e., the data used for training our model) to calculate the day-of-the-year precipitation climatological means which were subsequently used to calculate the anomalies for the verification period ($2003-2020$).

\subsection{Convection-Coupled Equatorial Waves}
Convection-coupled equatorial waves organize convective cells into ``synoptic scale'' disturbances in the tropics. Maintaining these waves in longer range model forecasts is important, because they are thought to play an important role in atmospheric predictability.
With the exception of the MJO, these waves are equatorially trapped wave solutions of the shallow-water equations on a beta-plane \citep{matsuno} and \citep{lindzen}. The Wheeler-Kiladis diagram \citep{Wheeler}, which can be obtained by performing a space-time spectral analysis of a scalar state variable tied to convective activity, has become a standard tool for the detection of convection-coupled equatorial waves in the atmosphere. It uses the dispersion relations of the waves as predicted by shallow-water theory, as well as the phenomenology based spectral characteristics in the case of the MJO, to detect the waves in two-dimensional fields of the scalar state variable. 

\begin{figure}
\includegraphics[scale=0.42]{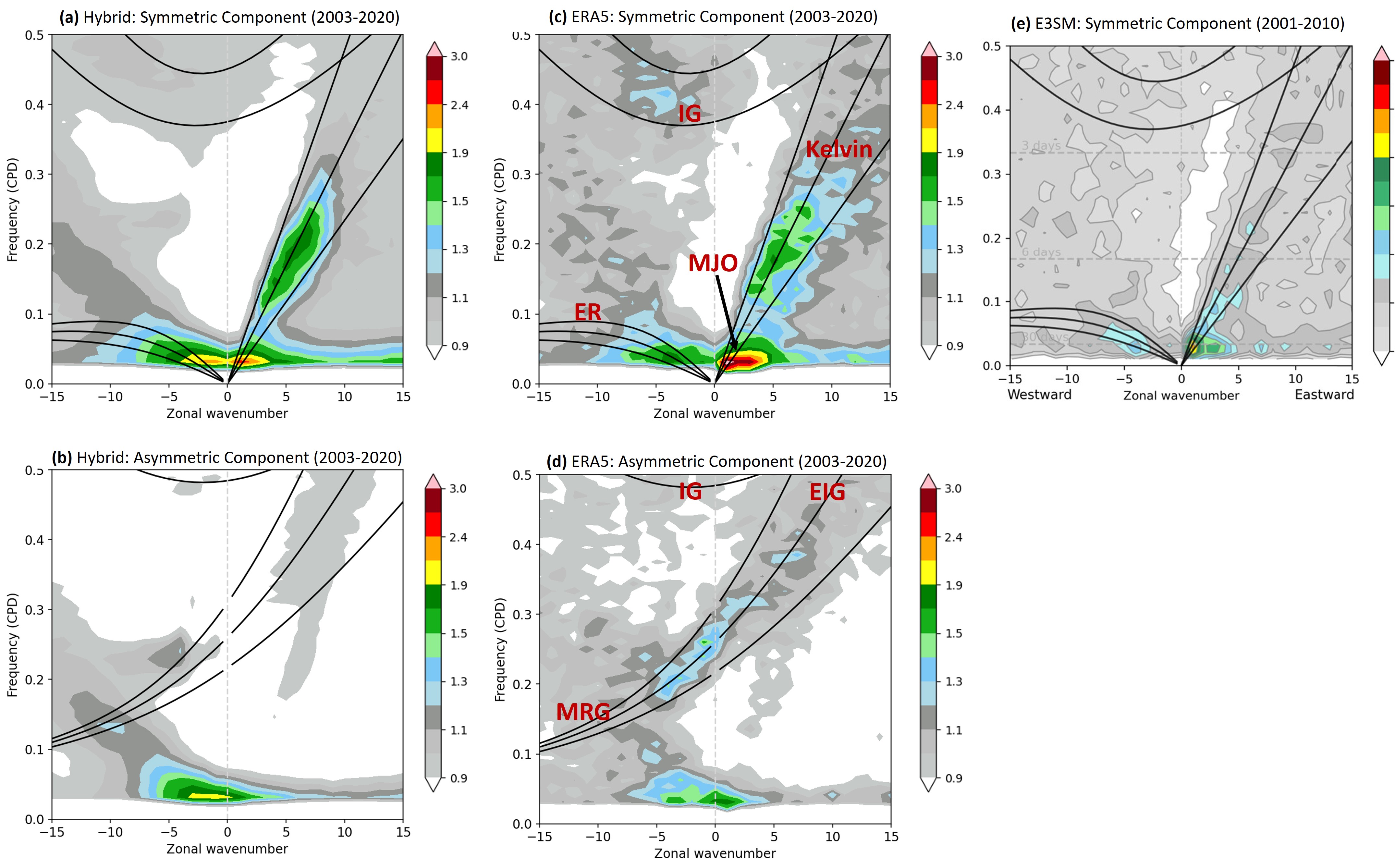}
\caption{The skill of the hybrid model forecasts in capturing the precipitation patterns associated with convection-coupled equatorial waves. The skill is illustrated with Wheeler-Kiladis diagrams. Shown is the power distribution of the precipitation in the frequency-wavenumber space for the (top) symmetric and (bottom) assymetric components of the waves for (left) the hybrid model forecasts (middle) ERA5 reanalyses and (right) E3SM model\citep[reproduced from][]{golaz}. The solid black lines indicate the dispersion relations from analytical equatorial wave solutions of the shallow-water equations for various values of the layer depth. The red labels indicate different types of wave solutions (ER corresponds to equatorial Rossby waves, IG to inertio-gravity waves, EIG to eastward inertio-gravity waves, MRG to mixed Rossby-gravity waves, and MJO to the Madden-Julian Oscillation)}
\label{fig:mjo_wk}
\end{figure}

Figure \ref{fig:mjo_wk} shows Wheeler-Kiladis diagrams for the 6\,h cumulative precipitation in the $15\degree$S - $15\degree$N latitude band. Panels~(a) and (b) show results for the hybrid model based on the 2-year long forecasts, while panels (c) and (d) show the corresponding diagrams for the ERA5 reanalyses. Panel~(e) shows the normalized symmetric component of a similar frequency-wavenumber analysis performed by \cite{golaz} for the E3SM model. The results indicate that the hybrid model captures the enhanced variability of the precipitation associated with Kelvin waves and equatorial Rossby waves. While our model has slightly lower spectral power in the frequency-wavenumber domain of the MJO (low frequencies and low positive zonal wave numbers), it has a power distribution similar to the reanalyses in this domain. In comparison, state-of-the-art conventional physics-based models such as E3SM and CFSv2 \citep[e.g.][]{goswami,golaz} struggle to capture these equatorial waves (e.g., panel (e)). Overall, we see that the tropical precipitation spectral power distribution from our model matches the observations well in many regimes, indicating that our model has learned a realistic representation of important aspects of the tropical convective processes.

\section{Conclusions and Outlook}\label{sec:Conclusion}
In this paper, we explored the potential of a hybrid modeling approach for weather prediction beyond the medium forecast range. We made several modifications to the previous (AEA23) version of our prototype hybrid model that implements this approach to further improve its forecast performance in this extended range. These modifications included adding a new ML-based internal prognostic variable to the oceanic component of the model for the heat content of the top 300 m layer of the ocean. Other modifications were made to the dynamics of the reservoir computing units of the ML-based oceanic model component to better handle the disparate time scales of the oceanic and atmospheric dynamics. The results of our forecast experiments showed that our model has useful forecast skill, similar to that of much higher resolution conventional purely physics-based models, for various oceanic and atmospheric phenomena that play important roles in predictability in the investigated forecast ranges. These phenomena include the El Ni\~no cycle and some of its extratropical teleconnections, and the signatures of convection-coupled equatorial waves in the precipitation fields. In particular, our model makes predictions of the El Ni\~no cycle that are more accurate for the first 4 months than those with the operational models of NMME. It also captures the strongest teleconnections patterns between El Ni\~no and the precipitation patterns around the world; and it is able to capture Kelvin wave and equatorial Rossby wave activity in 2-years long forecasts. We have obtained such results with a hybrid model that requires substantially more modest computing resources than the much higher resolution, currently operational, traditional physics-based models. 

The forecast performance of the model could most likely be further improved with the addition of more ML-based prognostic variables to better capture the two-way interactions between the atmosphere and components of the Earth system other than just the ocean, such as land and the cryosphere. Also, because the cumulative precipitation variable is purely ML-based, it could be trained on a processed observational product such as NASA's Integrated Multi-satellitE Retrievals for GPM (IMERG) data set rather than ERA5 reanalyses. This variable could even have higher spatial resolution than the other prognostic variables. These two changes could potentially greatly enhance the precipitation forecasts.

We believe that the forecast skill of our model, in terms of root-mean-square error, could be substantially increased by replacing our single-initial-condition deterministic forecasts with the mean of an ensemble of model forecasts. In fact, the ensemble approach is the standard operational approach to make model predictions beyond the medium range. Because ensemble averaging minimizes the forecast RMSE by filtering unpredictable details from the forecasts, as the forecast lead time increases, the ensemble mean converges to climatology \citep{Epstein,Leith}. At intermediate lead times, if the ensemble is initialized appropriately, the ensemble mean provides an optimal blend of the initial condition-dependent and the climatology-based information in the least-square sense. Because ML-based models are also typically trained to optimize forecasts in the least-square sense, an ML-based model solution with time step $\Delta t$ can be expected to resemble an ensemble-mean forecast at forecast lead time $\Delta t$. Hence, a forecast with our hybrid model is similar to a forecast that has been pieced together from ensemble mean forecasts of length $\Delta t$ for the atmosphere and length $\Delta t^\prime$ for the ocean. We could increase the time steps to minimize the root-mean-square errors at specific long forecast lead times, but that would lead to the loss of details at the shorter lead times. We speculate that the currently used time steps combined with the ensemble approach would provide a good balance between details and longer term performance. Furthermore, we note that a properly initialized and calibrated ensemble can be used for uncertainty quantification tasks, which is of significant importance at the subseasonal-to-seasonal time scales. As such, while we have not yet performed rigorous tests of our hybrid approach for ensemble forecasting, our computationally cheap approach can provide a promising avenue for such tasks.

We see two potential areas for the application of our hybrid modeling approach. First, it could be used for the hybridization of current operational models. Such a hybridization would have the potential to lead to substantial forecast improvements, which could otherwise, at best, be achieved only by years of incremental improvements of the purely physics-based models. Second, a hybrid model based on a low-resolution research AGCM like ours can be a powerful tool of academic research. The computing challenges involved are not overwhelming for a small academic research group in which the hands-on code development is primarily done by graduate research assistants. 

As a final comment, our results suggest the important point that the hybrid ML/physics-based modeling approach is potentially capable of skillfully modeling climatological phenomena whose spatial scales can be resolved at the resolution of the hybrid model (Figs. \ref{fig:enso_pcc_rmse}-\ref{fig:mjo_wk}; see also Figs. 2 and 3 of AEA23), without explicitly modeling the subgrid scale phenomena. The training implicitly includes the effects of unresolved scales on the resolved scales, because the training data, although itself essentially at the resolved scales, includes the real effects of the unresolved smaller scale dynamics in determining the observed larger scale dynamics that is being learned (this point has been previously emphasized by \cite{Wikner2020}). We note that this need not be the case for conventional physics-based models which typically employ subgrid scale modeling. While the parameters associated with the subgrid scale modeling can be tuned/calibrated, the accuracy of the subgrid scale model itself strongly affects the behavior of scales resolved by the model, and can be problematic, especially for a low resolution model.

\appendix

\section{ML Model Architecture, Training, and Inference}\label{sec: ML details}
\subsection{The Global and Local State Vectors}\label{sec: Global and Local State Vectors}
For our hybrid atmospheric model, we adopt the spatial discretization of the global domain that is used by the physics-based numerical model SPEEDY. While SPEEDY is a spectral transform model, it uses a discretized spatial grid with horizontal grid spacing of $3.75\degree \times 3.75\degree$ (so $n_h = 96 \times 48 = 36,864$ horizontal grid points (black dots in Fig. \ref{fig:domain_decomp})) to represent the input and output model state and to compute the nonlinear and parameterized terms of the physics-based prognostic equations. The SPEEDY model state consists of the two components of the three-dimensionally varying horizontal wind, the three-dimensionally varying temperature and specific humidity fields, and the two-dimensionally varying (natural logarithm of) surface pressure. The three-dimensionally varying state variables are defined at $n_v = 8$ vertical $\sigma$-levels (at $\sigma = 0.025, 0.095, 0.20, 0.34, 0.51, 0.685, 0.835,$ and $0.95$), where $\sigma$ is the ratio of pressure to the surface pressure. We add to the notation of the grid-based global atmospheric state vector $\mathbf{u}_G$ the super-script $P$ to represent the SPEEDY model state $\mathbf{u}^P_G$, the super-script $H$ to represent the hybrid atmospheric model state $\mathbf{u}^H_G$, and the super-script $a$ to represent the observation-based analysis (estimate) $\mathbf{u}^a_G$ of $\mathbf{u}_G$. In addition to the prognostic variables of SPEEDY, $\mathbf{u}^H_G$ can include additional prognostic variables. In our current version of the model, there is one such prognostic variable: $ln(10^3P+1)$, where $P$ is the $6$-h cumulative precipitation in units of meters. This prognostic variable is a purely ML-based state variable rather than a hybridized state variable based on the diagnostic precipitation variable of SPEEDY. The same transformed variable for the cumulative precipitation was used before in ML-based weather prediction models by \cite{Pathak2022} and \cite{Rasp}. We take $\mathbf{u}_G$ to be $\mathbf{u}_G^a$ during training and $\mathbf{u}_G^H$ during prediction. (See Fig. \ref{fig:arch_train_pred} and refer to Sec. \ref{sec: ML Model  Dynamics} for further details.)

The ML-only oceanic component of the model uses the same horizontal grid as the hybrid atmospheric component. The elements of the ocean model state vector are the grid values of the two-dimensionally varying fields of the SST and the heat content of the top 300 m layer of the ocean. We add the super-script $a$ to the notation of the grid-based global oceanic state vector $\mathbf{v}_G$, which represents the aforementioned two fields as a single vector, to denote its observation-based estimate $\mathbf{v}^a_G$. We add the super-script $M$ to $\mathbf{v}_G$ to denote the ML-only oceanic model state $\mathbf{v}_G^M$. We take $\mathbf{v}_G$ to be $\mathbf{v}_G^a$ during our model training, and $\mathbf{v}_G^M$ during prediction (see Fig. \ref{fig:arch_train_pred}). We note that heat content is an internal variable of the oceanic component of the model, because only the SST field serves as input to the atmospheric component.

We partition the global horizontal domain into $L=1,152$ sub-domains. Each sub-domain $l$ has a $7.5\degree \times 7.5\degree$ base of $4$ grid points (see the black dots enclosed by the blue box in Fig. \ref{fig:domain_decomp}) and the subdomains for the atmosphere include the $n_v$ $\sigma$-levels specified by SPEEDY. Let $\mathbf{u}_l$, $\mathbf{u}^P_l$ and $\mathbf{u}^H_l$ be the local sub-vectors of $\mathbf{u}_G$, $\mathbf{u}^P_G$ and $\mathbf{u}^H_G$, respectively, which consist of the elements of the related global state vectors that represent the atmospheric state in sub-domain $l$. Likewise, let $\mathbf{v}_l$ and   $\mathbf{v}_l^M$ be the sub-vectors of $\mathbf{v}_G$ and $\mathbf{v}^M_G$ that represent the oceanic state in sub-domain $l$. In addition, we define $\mathbf{u}^a_l$ and $\mathbf{v}^a_l$ to be the observation-based analyses (estimates) of $\mathbf{u}_l$ and $\mathbf{v}_l$. 

The different state variables can have different units and ranges (possibly separated by orders of magnitude), and for a given state variable the range can also vary with the geographic location and altitude. To avoid artificially under-emphasizing or over-emphasizing any particular state variable in the training of the ML components, at each grid point in a sub-domain $l$ we re-scale each state variable by subtracting the climatological mean and dividing by the climatological standard deviation of that state variable over all corresponding horizontal grid points at that altitude in sub-domain $l$ and over all time points in the training data. The re-scaled state variables are then used to form the local state vectors that are used as input to the ML-components of the model during training and when making predictions.  

\begin{figure}
\includegraphics[scale=0.4]{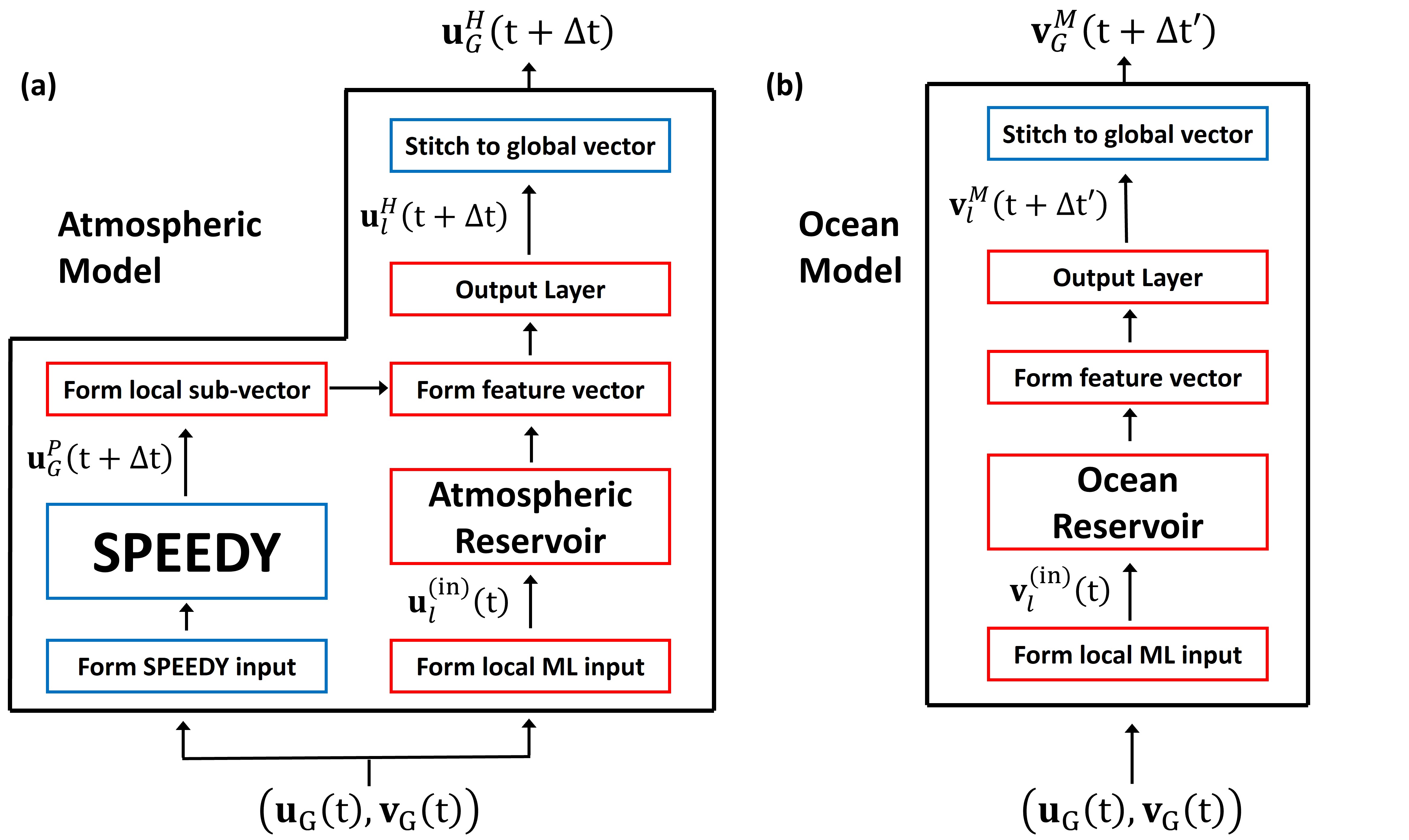}
\caption{Model architecture flow chart for the (a) hybrid atmospheric component, and (b) ML-only oceanic component. The notation is explained in detail in Sec. \ref{sec: Global and Local State Vectors} and Sec. \ref{sec: ML Model  Dynamics}. Red boxes represent operations that are performed locally (and in parallel) on each of the $L=1,152$ sub-domains. Blue boxes represent operations for which the outputs are global state vectors. Both model components receive $\mathbf{u}_G(t)$ and $\mathbf{v}_G(t)$ as input. The hybrid atmospheric (ML-only oceanic) component produces $\mathbf{u}_G^H(t+\Delta t)$ ($\mathbf{v}_G^M(t+\Delta t')$) as output, by stitching together local forecasts $\mathbf{u}_l^H(t+\Delta t)$ ($\mathbf{v}_l^M(t+\Delta t')$). In practice, the ocean model time step is larger than the atmospheric model time step, and for time steps where the ocean model is not advanced, the ocean model state is held constant.}
\label{fig:arch_diag}
\end{figure}

\subsection{ML Model Dynamics}\label{sec: ML Model  Dynamics}
The basic units of the ML-based components of the model are reservoir computers. \deleted{These units advance the ML-based component of the local atmospheric state vectors $\mathbf{u}^H_l(t)$ as well as the full local oceanic state vectors $\mathbf{v}^M_l(t)$ in time.} In principle, the reservoir computing (RC) time step can be different for the atmosphere and ocean (i.e., for $\mathbf{u}^H_l(t)$ and $\mathbf{v}^M_l(t)$). We choose the time step $\Delta t'$ for the ocean to be larger than the time step $\Delta t$ for the atmosphere by an integer multiple $n$ (i.e., $\Delta t' = n\Delta t$). When the oceanic component makes a $\Delta t'$-step forward forecast for time $t+\Delta t'$ starting at time $t$, the atmospheric component makes $n$ successive $\Delta t$-step forward forecasts starting at time $t$. During the $n$ $\Delta t$-step forward atmospheric forecasts, the ocean model state is updated on the first step and held constant thereafter; if $t$ is not a multiple of $\Delta t'$, then we set $\mathbf{v}_G(t)$ and $\mathbf{v}_l(t)$ to be equal to their values at the next multiple of $\Delta t'$. (We have experimented only with values of $\Delta t$ that have been substantially larger than the numerical integration time step for the physics-based SPEEDY model.) To propagate information about the state between subdomains, the RC inputs at time $t$ for the prediction of $\mathbf{u}_l(t+\Delta t)$ and $\mathbf{v}_l(t+\Delta t')$ describe the states in a local region that is extended in the horizontal directions (see the red box in Fig. \ref{fig:domain_decomp}). 

In addition to the components of $\mathbf{u}_G(t)$ that describe the atmospheric state in the extended local region, the input vector $\mathbf{u}^{(in)}_l(t)$ (Fig. \ref{fig:arch_diag}(a)) for the prediction of $\mathbf{u}_l(t+\Delta t)$
also includes the SST components of $\mathbf{v}_G(t)$ from the extended local region, and additional prescribed components that represent the incoming solar radiation at the top of the atmosphere.
The reservoir computer for each sub-domain $l$ has an internal state vector of dimension $N$ that we denote $\mathbf{r}_l(t) = [r_1(t), r_2(t), ..., r_N(t)]^T$ (see the red box labeled ``Atmospheric Reservoir" in Fig. \ref{fig:arch_diag}). This reservoir state evolves in time according to the following recurrent equation, driven by the corresponding local $m$-dimensional input vector $\mathbf{u}^{(in)}_l(t)$ of that sub-domain,
\begin{equation}\label{eqn:res_eq}
    \mathbf{r}_l(t+\Delta t) = (\mathbf{I} - \mathbf{C}_l) \mathbf{r}_l(t) + \mathbf{C}_l \tanh(\mathbf{A}_l\mathbf{r}_l(t) + \mathbf{B}_l\mathbf{u}^{(in)}_l(t)).
\end{equation}
where $\mathbf{B}_l$ is a $N \times m$ sparse matrix with only one nonzero element per row, and such that each column has the same number of nonzero elements. The nonzero elements of $\mathbf{B}_l$ are randomly chosen from a uniform distribution over the interval $[-b, b]$. The matrix $\mathbf{A}_l$ is called the reservoir network adjacency matrix and is generated as a $N \times N$ sparse, random matrix with an average of $d$ nonzero matrix elements in each row and each column (we refer to $d$ as the ``nodal degree"). It is then re-scaled by dividing each element by the computed spectral radius of the generated matrix and then multiplying each element by the desired spectral radius $\rho$. (The spectral radius of $\mathbf{A}_l$ is the magnitude of its largest eigenvalue.) The matrix $\mathbf{I}$ is a $N \times N$ identity matrix, and the ``leaking rate matrix" $\mathbf{C}_l = diag(c_1,c_2, ..., c_N)$ is a diagonal matrix of leaking rates that determine the amount of memory retained by the current reservoir state of the past reservoir states. The ``activation function" $\tanh(\cdot)$ is an $N$-vector formed from the $N$ scalars obtained by the $\tanh$ operating element-wise on its $N$-vector argument.

Meanwhile, the global predicted state $\mathbf{u}^P_G(t + \Delta t)$ is obtained by forecasting with SPEEDY for time $\Delta t$, using $\mathbf{u}_G(t)$ as initial condition and the SST components of $\mathbf{v}_G(t)$ as boundary condition. The local sub-vector $\mathbf{u}^P_l(t + \Delta t)$ is combined with local reservoir state $\mathbf{r}_l(t + \Delta t)$ to form the augmented atmospheric model reservoir state $\mathbf{\Tilde{r}}_l(t + \Delta t) = [\mathbf{u}^P_l(t + \Delta t), \chi(\mathbf{r}_l(t + \Delta t))]^T$, where $\chi(\cdot)$ squares every other element of its vector-valued argument (e.g., for $\mathbf{x} = [x_1, x_2, x_3, x_4, ...]^T$, $\chi(\mathbf{x}) = [x_1, x^2_2, x_3, x^2_4, ...]^T$) \citep[][]{patel,pathakPRL}. The augmented state is linearly mapped to the local hybrid atmospheric model state according to 
\begin{equation}\label{eqn:output_atmo}
    \mathbf{u}^H_l(t + \Delta t) = \mathbf{W}_l \mathbf{\Tilde{r}}_l(t + \Delta t) = \mathbf{W}^P_l\mathbf{u}^P_l(t + \Delta t) + \mathbf{W}^R_l\chi(\mathbf{r}_l(t+\Delta t)), 
\end{equation}
where $\mathbf{W}_l = [\mathbf{W}^P_l\quad \mathbf{W}^R_l]$ is an output weight matrix of trainable parameters. Additionally, for the hybrid atmospheric model $\mathbf{W}_l = \left[ \mathbf{W}^P_l\quad \mathbf{W}^R_l \right]$. $\mathbf{u}^P_l(t + \Delta t)$ is obtained from $\mathbf{u}^P_G(t + \Delta t)$, which is obtained by making a $\Delta t$ time step forecast with SPEEDY using $\mathbf{u}^P_G(t)$ as the initial condition. 

For the ML-only ocean component, in addition to the components of $\mathbf{v}_G(t)$ that describe the oceanic state in the extended local region, the input vector $\mathbf{v}^{(in)}_l(t)$ (Fig. \ref{fig:arch_diag}(b)) for the prediction of $\mathbf{v}_l(t+\Delta t')$ also includes the components of $\mathbf{u}_G(t)$ corresponding to the lowest atmospheric model level from the extended local region and additional prescribed components that represent the incoming solar radiation at the top of the atmosphere. (We implement a simple land/ocean mask in the following way. Only sub-domains that contain at least one grid point located over an ocean are assigned a local ocean reservoir for predicting the local oceanic state. If the sub-domain contains a mix of land and ocean grid points, the elements of the local oceanic state vector corresponding to the ``SST" values at land grid points are set to $272.0$K and the elements of the local oceanic state vector corresponding to the ``top 300m ocean heat content" at land grid points are set to $0$ Jm$^{-2}$, during both training and prediction. Specifically, for prediction, the local reservoir for such a sub-domain does predict the ``SST"  and ``top 300m ocean heat content" values at the land grid points, but this prediction is overwritten by setting that value to $272.0$K for the SST and to $0$ Jm$^{-2}$ for the top 300m ocean heat content.) The local reservoir state vector $\mathbf{r}'_l(t)$ of the ocean reservoir corresponding to sub-domain $l$ (see the red box labeled ``Ocean Reservoir" in Fig. \ref{fig:arch_diag}(b)) evolves according to the following recurrent equation, driven by the corresponding local input vector $\mathbf{v}^{(in)}_l(t)$ of that sub-domain,
\begin{equation}\label{eqn:res_eq_ocean}
    \mathbf{r}'_l(t+\Delta t') = (\mathbf{I} - \mathbf{C}'_l) \mathbf{r}'_l(t) + \mathbf{C}'_l \tanh(\mathbf{A}'_l\mathbf{r}'_l(t) + \mathbf{B}'_l\mathbf{v}^{(in)}_l(t)),
\end{equation}
where the prime notation is used to emphasize that the ML components of the hybrid atmospheric prognostic equations and the ML-only oceanic prognostic equations use independent reservoirs (see the separate branches for computations for the atmospheric and oceanic models with separate ``Atmospheric Reservoir" and ``Ocean Reservoir" in Fig. \ref{fig:arch_diag}). The local oceanic model state $\mathbf{v}^M_l(t+\Delta t')$ is then calculated as 
\begin{equation}\label{eqn:output_ocean}
\mathbf{v}^M_l(t+\Delta t') = \mathbf{W}'_l \mathbf{\Tilde{r}'}_l(t + \Delta t') = \mathbf{W}'_l \chi(\mathbf{r}'_l(t+\Delta t')).
\end{equation}

Equations (\ref{eqn:res_eq_ocean}) and (\ref{eqn:output_ocean}) are applied only when $t$ is a multiple $\Delta t' = n\Delta t$. In order that $\mathbf{v}^M_l(t)$ be defined at every time step of the atmospheric model component, we define $\mathbf{v}^M_l(t + j\Delta t) = \mathbf{v}^M_l(t + n\Delta t)$ for $j=1,2,..., n-1$ whenever $t$ is a multiple of $\Delta t'$. This represents a particularly simple form of the interpolation depicted in Fig. \ref{fig:arch_train_pred}(b). The parameters of the output weight matrices $\mathbf{W}_l$ and $\mathbf{W}'_l$   are learned from reanalysis data (see Sec. \ref{sec:Data}) during training, the details of which are outlined in the next subsection.

\subsection{Model Training}\label{sec: Model Training}
For a sub-domain $l$ of the atmospheric model, the objective of the ML training is as follows. Given target training data $\mathbf{u}^a_l(k \Delta t)$ at times $k\Delta t$ for $k = -K, -K+1, -K+2, ..., 0$, determine parameters $\mathbf{W}_l$ for which the one-step forward model forecasts best fit the training data in a least-squares sense. This is done by minimizing the following cost function,
\begin{equation}\label{eqn:cost_func}
    J(\mathbf{W}_l) = \sum_{k=-K+K_T}^{0} \lVert \mathbf{W}_l \mathbf{\Tilde{r}}_l(k\Delta t) - \mathbf{u}^a_l(k\Delta t) \rVert^2 + \beta^P\lVert \mathbf{W}^P_l \rVert^2 +
    \beta^R\lVert \mathbf{W}^R_l \rVert^2,
\end{equation}
where the adjustable Tikhonov regularization parameter $\beta$ is included to prevent over-fitting \citep[][]{Tikhonov}, and $K_T$ denotes the number of samples that are discarded as transients. Note, the reservoir computers of all sub-domains are trained independently in an identical manner (see the blue box in Fig. \ref{fig:arch_diag}). For the hybrid prognostic equations, the solution to the minimization problem is
\begin{equation}\label{eqn:wout_atmo}
    \mathbf{W}_l = \left( \mathbf{U}^a_l (\mathbf{U}^P_l)^T\qquad \mathbf{U}^a_l\mathbf{R}^T \right) \begin{bmatrix}
        \mathbf{U}^P_l(\mathbf{U}^P_l)^T + \beta^P\mathbf{I}\qquad \mathbf{U}^P_l\mathbf{R}^T \\
        \mathbf{R}(\mathbf{U}^P_l)^T\quad \mathbf{R}\mathbf{R}^T + \beta^R\mathbf{I} 
        \end{bmatrix}^{-1}
\end{equation}
where column $k$ of matrix $\mathbf{U}^a_l$ is $\mathbf{u}^a_l(k\Delta t)$, column $k$ of the matrix $\mathbf{U}^P_l$ is $\mathbf{u}^P_l(k\Delta t)$, and $\beta^P$ and $\beta^R$ are the adjustable regularization parameters for the physics-based model output and reservoir state output, respectively. The column $k$ of the matrix $\mathbf{R}$ is $\chi(\mathbf{r}(k\Delta t))$, and the reservoir states $\mathbf{r}(k\Delta t)$ for $k=-K+K_T,-K+K_T+1, ..., 0$ are obtained by iterating Eq. (\ref{eqn:res_eq}).

For the ML-only oceanic prognostic equations, the solution to the minimization problem for the equivalent of the cost function of Eq.~(\ref{eqn:cost_func}) for the ocean is
\begin{equation}\label{eqn:wout_ocean}
    \mathbf{W}'_l = \mathbf{V}^a (\mathbf{R}')^T \left(\mathbf{R}'(\mathbf{R}')^T + \beta'\mathbf{I} \right)^{-1}
\end{equation}
where column $k$ of matrix $\mathbf{V}^a_l$ is $\mathbf{v}^a_l(k\Delta t')$, column $k$ of $\mathbf{R}'$ is $\chi(\mathbf{r}'(k\Delta t'))$, and $\beta'$ is the Tikhonov regularization parameter.

\subsection{Synchronization and Prediction}\label{sec:sync_and_predict}
In order to make a forecast with the trained model from a forecast start time $K_f\Delta t$, we need both the observation-based data and the \emph{synchronized} reservoir states to be available at that time for all sub-domains. The synchronized reservoir state $\mathbf{r}_l(K_f\Delta t)$ at time $K_f\Delta t$ for sub-domain $l$ can be obtained by iterating Eq. (\ref{eqn:res_eq}) using the appropriate observation-based data for times $k\Delta t$ where $k = K_f-K_T, K_f-K_T-1,K_f-K_T-2, ..., K_f-1$. We allow $K_T$ to be large enough such that, to a high degree of accuracy, $\mathbf{r}_l(K_f\Delta t)$ does not depend on the arbitrary choice of initialization of the reservoir state $\mathbf{r}_l((K_f-K_T)\Delta t)$ (this is what we mean by ``synchronization"). The reservoir state $\mathbf{r}_l((K_f+1)\Delta t)$ at time $(K_f+1)\Delta t$ can then be obtained by using $\mathbf{r}_l(K_f\Delta t)$ and the corresponding observation-based data according to Eq. (\ref{eqn:res_eq}). The property that the reservoir states do not depend on their arbitrary initial choice is called the ``echo state property" \citep[][]{Jaeger,Luk,Luk+Jaeger}, and we assume that relevant hyperparameters (e.g., the spectral radius, $\rho$, of $\mathbf{A}_l$) of the reservoir are chosen so that the conditions for this property are satisfied. We can then use the appropriate prediction equations (Eq. (\ref{eqn:output_atmo}) or Eq. (\ref{eqn:output_ocean})) to make a $\Delta t$-step forward prediction. The forecasts of all sub-domains, which can be produced independently and in parallel, are pieced together to create the corresponding global forecast ($\mathbf{u}^H_G((K_f+1)\Delta t)$ and $\mathbf{v}^M_G((K_f+1)\Delta t)$) at time $(K_f+1)\Delta t$ (see Fig. \ref{fig:arch_diag}). A $2\Delta t$-step forward prediction can be obtained by advancing the local reservoir states forward using Eq. (\ref{eqn:res_eq}) and by setting $\mathbf{u}^{(in)}_l((K_f+1)\Delta t)$ or $\mathbf{v}^{(in)}_l((K_f+1)\Delta t)$ equal to the appropriate sub-vector of the $(K_f+1)\Delta t$ global forecast vector corresponding to the sub-domain $l$. The updated reservoir state $\mathbf{r}_l((K_f+2)\Delta t)$ is then used to make a forecast for time $(K_f+2)\Delta t$. In this way, we can operate the model in an auto-regressive configuration where the model output at time $\Delta t$ is fed back in as input to the model when making a forecast for time $2\Delta t$. This allows us to make forecasts of arbitrary lengths into the future.

\subsection{Choice of Hyperparameters}\label{sec: ML_HP}
Table \ref{tab:ocean_HP} lists the hyperparameter values for the set of hyperparameters of the ML-only oceanic model. See Sec. \ref{sec:Exp Design} for discussion of values chosen.

\begin{table}
\caption{Hyperparameters of the ML-only ocean model.}
\label{tab:ocean_HP}
\centering
\begin{tabular}{l c}
\hline
Hyperparameter  & Value  \\
\hline
$N'$  & 6000   \\ 
$d'$  & 6   \\ 
$\rho'$  & 0.9   \\
$b'$  & 0.6   \\
$\beta'$  & 100   \\
$c'_i$  & $[10^{-3},1]$   \\
$\Delta t'$  & 72 hours   \\
\hline
\end{tabular}
\end{table}

\section{Forecast Verification Metrics}\label{subsec:metrics}
\subsection{Pearson Correlation Coefficient}\label{subsubsec:PCC}
Letting $t$ denote the prediction lead time, we use $\{x_i(t)\}_{i=1,2,...,N_T}$ for a collection of $N_T$ predicted uni-variate time series and $\{y_i(t)\}_{i=1,2,...,N_T}$ for the corresponding collection of $N_T$ ``ground truth" uni-variate time series. The Pearson correlation coefficient as a function of prediction lead time $t$ is given by
\begin{equation}\label{eqn:PCC}
    r(t) = \frac{\sum_{i \in S} (x_i(t) - \bar{x}(t))(y_i(t) - \bar{y}(t))}{\sqrt{\sum_{i \in S} (x_i(t) - \bar{x}(t))^2 \sum_{i \in S}(y_i(t) - \bar{y}(t))^2}},
\end{equation}
where $S$ is the set of indices denoting the forecasts used when computing the PCC (e.g., we may only include forecasts with start dates in the month of June when assessing the forecast skill of the model for the June-July-August season). The over-bar $\bar{(\ast)}$ denotes the average of $(\ast)_i$ for $i \in S$. When considering the ability of a model to predict an anomaly signal, it is a common practice in the literature to accept as a valid forecast horizon the maximum prediction lead time for which the PCC is greater than or equal to $0.5$ (i.e., $\max \{t \vert r(t) \geq 0.5\}$) \citep[][]{Owens,Ham,Zhou}. We follow this convention as one of our metrics to assess forecast skill. We also calculate the $95\%$ confidence interval associated with our PCC values to estimate their statistical significance. 

\subsection{Root-Mean-Square Error}\label{subsubsec:RMSE}
The root-mean-square error as a function of lead time $t$ averaged over a collection of univariate time series is given by
\begin{equation}\label{eqn:RMSE}
    RMSE(t) = \sqrt{\frac{\sum_{i \in S} (x_i(t) - y_i(t))^2}{\vert S \vert}}
\end{equation}
where $\vert S \vert$ denotes the number of elements in $S$. When appropriate, we also report the RMSE for the persistence and climatology based forecast. The relationship between model forecast errors and those of the persistence and climatology is well understood, (e.g., Section 3.8 of \cite{Szunyogh}). The RMSE of the persistence based forecast, while smaller than the RMSE of the climatology based forecast at shorter lead times, saturates at a value $\sqrt{2}$ times greater than the climatology error at longer lead times. We define our model to have useful forecast skill if it produces forecasts with a lower value of the RMSE than that of both the persistence based forecasts and the climatologically based forecasts. If our model has realistic climatology, then the RMSE of the forecasts at longer lead times should saturate at the same level as that for the persistence based forecasts.

\section*{Open Research}
The weights of the trained hybrid model used in this paper are available on Zenodo via \url{https://doi.org/10.5281/zenodo.11390866} \citep[][]{dhruvitweights}. Sample data in the format expected by the model is available on Zenodo via \url{https://doi.org/10.5281/zenodo.14043079} \citep[][]{dhruvitdata} The code to perform predictions with the trained model (as well as the code to train a new model from scratch) and reproduce the results of this paper is available on Github-Zenodo via \url{https://doi.org/10.5281/zenodo.14038114} \citep[][]{dhruvitcode}. Data of the NMME models' hindcast of the sea-surface temperatures was obtained from the NMME System Phase II data repository \citep[][]{Kirtman}, available at \url{https://www.earthsystemgrid.org/search.html?Project=NMME}. The figures in this paper were prepared using Matplotlib version 3.7.2 \citep[][]{Hunter:2007}, available under the Matplotlib license at \url{https://matplotlib.org/}, and Cartopy version 0.21.1 \citep[][]{Cartopy}, available under the Open Government License at \url{https://scitools.org.uk/cartopy/docs/latest/index.html}. The analysis in this paper was performed using Xarray version 2023.7.0, Numpy 1.24.4 \citep[][]{harris2020array}, available under the Numpy license at \url{https://numpy.org/doc/2.1/index.html}, and Scipy version 1.11.1 \citep[][]{2020SciPy-NMeth}.



\acknowledgments
The work of Istvan Szunyogh, Brian Hunt, and Edward Ott on this research was partly supported by the Office of Naval Research under grant N00014-22-1-2319. Portions of this research were conducted with the advanced computing resources provided by Texas A\& M High Performance Research Computing. Troy Arcomano was partly supported by the Global Change Fellowship in the Environmental Science Division at Argonne National Laboratory (grant no. LDRD 2023-0236). We acknowledge the agencies that support the NMME-Phase II system, and we thank the climate modeling groups (Environment Canada, NASA, NCAR, NOAA/GFDL, NOAA/NCEP, and University of Miami) for producing and making available their model output. NOAA/NCEP, NOAA/CTB, and NOAA/CPO jointly provided coordinating support and led development of the NMME-Phase II system. We would also like to thank the reviewers for their time in reviewing our work prior to publication and for providing important and useful feedback. 

\bibliography{bibliography}

\begin{thebibliography}{}

\bibitem [\protect \citeauthoryear {%
Alexander%
\ \protect \BOthers {.}}{%
Alexander%
\ \protect \BOthers {.}}{%
{\protect \APACyear {2002}}%
}]{%
Alexander}
\APACinsertmetastar {%
Alexander}%
\begin{APACrefauthors}%
Alexander, M\BPBI A.%
, Bladé, I.%
, Newman, M.%
, Lanzante, J\BPBI R.%
, Lau, N\BHBI C.%
\BCBL {}\ \BBA {} Scott, J\BPBI D.%
\end{APACrefauthors}%
\unskip\
\newblock
\APACrefYearMonthDay{2002}{}{}.
\newblock
{\BBOQ}\APACrefatitle {The Atmospheric Bridge: The Influence of ENSO Teleconnections on Air–Sea Interaction over the Global Oceans} {The atmospheric bridge: The influence of enso teleconnections on air–sea interaction over the global oceans}.{\BBCQ}
\newblock
\APACjournalVolNumPages{Journal of Climate}{15}{16}{2205 - 2231}.
\newblock
\begin{APACrefDOI} \doi{10.1175/1520-0442(2002)015<2205:TABTIO>2.0.CO;2} \end{APACrefDOI}
\PrintBackRefs{\CurrentBib}

\bibitem [\protect \citeauthoryear {%
Arcomano%
, Szunyogh%
, Wikner%
, Hunt%
\BCBL {}\ \BBA {} Ott%
}{%
Arcomano%
\ \protect \BOthers {.}}{%
{\protect \APACyear {2023}}%
}]{%
Arcomano2023}
\APACinsertmetastar {%
Arcomano2023}%
\begin{APACrefauthors}%
Arcomano, T.%
, Szunyogh, I.%
, Wikner, A.%
, Hunt, B\BPBI R.%
\BCBL {}\ \BBA {} Ott, E.%
\end{APACrefauthors}%
\unskip\
\newblock
\APACrefYearMonthDay{2023}{}{}.
\newblock
{\BBOQ}\APACrefatitle {A Hybrid Atmospheric Model Incorporating Machine Learning Can Capture Dynamical Processes Not Captured by Its Physics-Based Component} {A hybrid atmospheric model incorporating machine learning can capture dynamical processes not captured by its physics-based component}.{\BBCQ}
\newblock
\APACjournalVolNumPages{Geophysical Research Letters}{50}{8}{e2022GL102649}.
\newblock
\begin{APACrefDOI} \doi{https://doi.org/10.1029/2022GL102649} \end{APACrefDOI}
\PrintBackRefs{\CurrentBib}

\bibitem [\protect \citeauthoryear {%
Arcomano%
\ \protect \BOthers {.}}{%
Arcomano%
\ \protect \BOthers {.}}{%
{\protect \APACyear {2022}}%
}]{%
Arcomano_2022}
\APACinsertmetastar {%
Arcomano_2022}%
\begin{APACrefauthors}%
Arcomano, T.%
, Szunyogh, I.%
, Wikner, A.%
, Pathak, J.%
, Hunt, B\BPBI R.%
\BCBL {}\ \BBA {} Ott, E.%
\end{APACrefauthors}%
\unskip\
\newblock
\APACrefYearMonthDay{2022}{}{}.
\newblock
{\BBOQ}\APACrefatitle {A Hybrid Approach to Atmospheric Modeling That Combines Machine Learning With a Physics-Based Numerical Model} {A hybrid approach to atmospheric modeling that combines machine learning with a physics-based numerical model}.{\BBCQ}
\newblock
\APACjournalVolNumPages{Journal of Advances in Modeling Earth Systems}{14}{3}{e2021MS002712}.
\newblock
\begin{APACrefDOI} \doi{https://doi.org/10.1029/2021MS002712} \end{APACrefDOI}
\PrintBackRefs{\CurrentBib}

\bibitem [\protect \citeauthoryear {%
Bi%
\ \protect \BOthers {.}}{%
Bi%
\ \protect \BOthers {.}}{%
{\protect \APACyear {2023}}%
}]{%
Bi2023}
\APACinsertmetastar {%
Bi2023}%
\begin{APACrefauthors}%
Bi, K.%
, Xie, L.%
, Zhang, H.%
, Chen, X.%
, Gu, X.%
\BCBL {}\ \BBA {} Tian, Q.%
\end{APACrefauthors}%
\unskip\
\newblock
\APACrefYearMonthDay{2023}{}{}.
\newblock
{\BBOQ}\APACrefatitle {Accurate medium-range global weather forecasting with 3D neural networks} {Accurate medium-range global weather forecasting with 3d neural networks}.{\BBCQ}
\newblock
\APACjournalVolNumPages{Nature}{619}{7970}{533--538}.
\newblock
\begin{APACrefDOI} \doi{10.1038/s41586-023-06185-3} \end{APACrefDOI}
\PrintBackRefs{\CurrentBib}

\bibitem [\protect \citeauthoryear {%
Bjerknes%
}{%
Bjerknes%
}{%
{\protect \APACyear {1969}}%
}]{%
Bjerknes}
\APACinsertmetastar {%
Bjerknes}%
\begin{APACrefauthors}%
Bjerknes, J.%
\end{APACrefauthors}%
\unskip\
\newblock
\APACrefYearMonthDay{1969}{}{}.
\newblock
{\BBOQ}\APACrefatitle {ATMOSPHERIC TELECONNECTIONS FROM THE EQUATORIAL PACIFIC} {Atmospheric teleconnections from the equatorial pacific}.{\BBCQ}
\newblock
\APACjournalVolNumPages{Monthly Weather Review}{97}{3}{163 - 172}.
\newblock
\begin{APACrefDOI} \doi{10.1175/1520-0493(1969)097<0163:ATFTEP>2.3.CO;2} \end{APACrefDOI}
\PrintBackRefs{\CurrentBib}

\bibitem [\protect \citeauthoryear {%
Brassington%
\ \protect \BOthers {.}}{%
Brassington%
\ \protect \BOthers {.}}{%
{\protect \APACyear {2015}}%
}]{%
Brassington2015}
\APACinsertmetastar {%
Brassington2015}%
\begin{APACrefauthors}%
Brassington, G\BPBI B.%
, Martin, M\BPBI J.%
, Tolman, H\BPBI L.%
, Akella, S.%
, Balmeseda, M.%
, Chambers, C\BPBI R\BPBI S.%
\BDBL {}Todling, R.%
\end{APACrefauthors}%
\unskip\
\newblock
\APACrefYearMonthDay{2015}{}{}.
\newblock
{\BBOQ}\APACrefatitle {Progress and challenges in short- to medium-range coupled prediction} {Progress and challenges in short- to medium-range coupled prediction}.{\BBCQ}
\newblock
\APACjournalVolNumPages{Journal of Operational Oceanography}{8}{sup2}{s239--s258}.
\newblock
\begin{APACrefDOI} \doi{10.1080/1755876X.2015.1049875} \end{APACrefDOI}
\PrintBackRefs{\CurrentBib}

\bibitem [\protect \citeauthoryear {%
Chattopadhyay%
\ \BBA {} Hassanzadeh%
}{%
Chattopadhyay%
\ \BBA {} Hassanzadeh%
}{%
{\protect \APACyear {2023}}%
}]{%
chattopadhyay2023longterm}
\APACinsertmetastar {%
chattopadhyay2023longterm}%
\begin{APACrefauthors}%
Chattopadhyay, A.%
\BCBT {}\ \BBA {} Hassanzadeh, P.%
\end{APACrefauthors}%
\unskip\
\newblock
\APACrefYearMonthDay{2023}{}{}.
\newblock
\APACrefbtitle {Long-term instabilities of deep learning-based digital twins of the climate system: The cause and a solution.} {Long-term instabilities of deep learning-based digital twins of the climate system: The cause and a solution.}
\newblock
\begin{APACrefDOI} \doi{https://doi.org/10.48550/arXiv.2304.07029} \end{APACrefDOI}
\PrintBackRefs{\CurrentBib}

\bibitem [\protect \citeauthoryear {%
K.~Chen%
\ \protect \BOthers {.}}{%
K.~Chen%
\ \protect \BOthers {.}}{%
{\protect \APACyear {2023}}%
}]{%
chen2023fengwu}
\APACinsertmetastar {%
chen2023fengwu}%
\begin{APACrefauthors}%
Chen, K.%
, Han, T.%
, Gong, J.%
, Bai, L.%
, Ling, F.%
, Luo, J\BHBI J.%
\BDBL {}Ouyang, W.%
\end{APACrefauthors}%
\unskip\
\newblock
\APACrefYearMonthDay{2023}{}{}.
\newblock
\APACrefbtitle {FengWu: Pushing the Skillful Global Medium-range Weather Forecast beyond 10 Days Lead.} {Fengwu: Pushing the skillful global medium-range weather forecast beyond 10 days lead.}
\newblock
\begin{APACrefDOI} \doi{https://doi.org/10.48550/arXiv.2304.02948} \end{APACrefDOI}
\PrintBackRefs{\CurrentBib}

\bibitem [\protect \citeauthoryear {%
L.~Chen%
, Zhong%
, Wu%
\BCBL {}\ \protect \BOthers {.}}{%
L.~Chen%
, Zhong%
, Wu%
\BCBL {}\ \protect \BOthers {.}}{%
{\protect \APACyear {2023}}%
}]{%
chen2023fuxis2s}
\APACinsertmetastar {%
chen2023fuxis2s}%
\begin{APACrefauthors}%
Chen, L.%
, Zhong, X.%
, Wu, J.%
, Chen, D.%
, Xie, S.%
, Chao, Q.%
\BDBL {}Qi, Y.%
\end{APACrefauthors}%
\unskip\
\newblock
\APACrefYearMonthDay{2023}{}{}.
\newblock
\APACrefbtitle {FuXi-S2S: An accurate machine learning model for global subseasonal forecasts.} {Fuxi-s2s: An accurate machine learning model for global subseasonal forecasts.}
\newblock
\begin{APACrefDOI} \doi{https://doi.org/10.48550/arXiv.2312.09926} \end{APACrefDOI}
\PrintBackRefs{\CurrentBib}

\bibitem [\protect \citeauthoryear {%
L.~Chen%
, Zhong%
, Zhang%
\BCBL {}\ \protect \BOthers {.}}{%
L.~Chen%
, Zhong%
, Zhang%
\BCBL {}\ \protect \BOthers {.}}{%
{\protect \APACyear {2023}}%
}]{%
chen2023fuxi}
\APACinsertmetastar {%
chen2023fuxi}%
\begin{APACrefauthors}%
Chen, L.%
, Zhong, X.%
, Zhang, F.%
, Cheng, Y.%
, Xu, Y.%
, Qi, Y.%
\BCBL {}\ \BBA {} Li, H.%
\end{APACrefauthors}%
\unskip\
\newblock
\APACrefYearMonthDay{2023}{}{}.
\newblock
\APACrefbtitle {FuXi: A cascade machine learning forecasting system for 15-day global weather forecast.} {Fuxi: A cascade machine learning forecasting system for 15-day global weather forecast.}
\newblock
\begin{APACrefDOI} \doi{https://doi.org/10.48550/arXiv.2306.12873} \end{APACrefDOI}
\PrintBackRefs{\CurrentBib}

\bibitem [\protect \citeauthoryear {%
Clarke%
\ \BBA {} Van~Gorder%
}{%
Clarke%
\ \BBA {} Van~Gorder%
}{%
{\protect \APACyear {2003}}%
}]{%
Clarke}
\APACinsertmetastar {%
Clarke}%
\begin{APACrefauthors}%
Clarke, A\BPBI J.%
\BCBT {}\ \BBA {} Van~Gorder, S.%
\end{APACrefauthors}%
\unskip\
\newblock
\APACrefYearMonthDay{2003}{}{}.
\newblock
{\BBOQ}\APACrefatitle {Improving El Niño prediction using a space-time integration of Indo-Pacific winds and equatorial Pacific upper ocean heat content} {Improving el niño prediction using a space-time integration of indo-pacific winds and equatorial pacific upper ocean heat content}.{\BBCQ}
\newblock
\APACjournalVolNumPages{Geophysical Research Letters}{30}{7}{}.
\newblock
\begin{APACrefDOI} \doi{https://doi.org/10.1029/2002GL016673} \end{APACrefDOI}
\PrintBackRefs{\CurrentBib}

\bibitem [\protect \citeauthoryear {%
Dai%
\ \BBA {} Wigley%
}{%
Dai%
\ \BBA {} Wigley%
}{%
{\protect \APACyear {2000}}%
}]{%
Dai}
\APACinsertmetastar {%
Dai}%
\begin{APACrefauthors}%
Dai, A.%
\BCBT {}\ \BBA {} Wigley, T\BPBI M\BPBI L.%
\end{APACrefauthors}%
\unskip\
\newblock
\APACrefYearMonthDay{2000}{}{}.
\newblock
{\BBOQ}\APACrefatitle {Global patterns of ENSO-induced precipitation} {Global patterns of enso-induced precipitation}.{\BBCQ}
\newblock
\APACjournalVolNumPages{Geophysical Research Letters}{27}{9}{1283-1286}.
\newblock
\begin{APACrefDOI} \doi{https://doi.org/10.1029/1999GL011140} \end{APACrefDOI}
\PrintBackRefs{\CurrentBib}

\bibitem [\protect \citeauthoryear {%
de Andrade%
, Coelho%
\BCBL {}\ \BBA {} Cavalcanti%
}{%
de Andrade%
\ \protect \BOthers {.}}{%
{\protect \APACyear {2019}}%
}]{%
Andrade}
\APACinsertmetastar {%
Andrade}%
\begin{APACrefauthors}%
de Andrade, F.%
, Coelho, C.%
\BCBL {}\ \BBA {} Cavalcanti, I.%
\end{APACrefauthors}%
\unskip\
\newblock
\APACrefYearMonthDay{2019}{}{}.
\newblock
{\BBOQ}\APACrefatitle {Global preiciptation hindcast quality assessment of the Subseasonal to Seasonal (S2S) prediction project models} {Global preiciptation hindcast quality assessment of the subseasonal to seasonal (s2s) prediction project models}.{\BBCQ}
\newblock
\APACjournalVolNumPages{Climate Dynamics}{52}{}{5451-5472}.
\newblock
\begin{APACrefDOI} \doi{https://doi.org/10.1007/s00382-018-4457-z} \end{APACrefDOI}
\PrintBackRefs{\CurrentBib}

\bibitem [\protect \citeauthoryear {%
Domeisen%
\ \protect \BOthers {.}}{%
Domeisen%
\ \protect \BOthers {.}}{%
{\protect \APACyear {2022}}%
}]{%
Domeisen}
\APACinsertmetastar {%
Domeisen}%
\begin{APACrefauthors}%
Domeisen, D\BPBI I\BPBI V.%
, White, C\BPBI J.%
, Afargan-Gerstman, H.%
, Ángel G.~Muñoz%
, Janiga, M\BPBI A.%
, Vitart, F.%
\BDBL {}Tian, D.%
\end{APACrefauthors}%
\unskip\
\newblock
\APACrefYearMonthDay{2022}{}{}.
\newblock
{\BBOQ}\APACrefatitle {Advances in the Subseasonal Prediction of Extreme Events: Relevant Case Studies across the Globe} {Advances in the subseasonal prediction of extreme events: Relevant case studies across the globe}.{\BBCQ}
\newblock
\APACjournalVolNumPages{Bulletin of the American Meteorological Society}{103}{6}{E1473 - E1501}.
\newblock
\begin{APACrefDOI} \doi{10.1175/BAMS-D-20-0221.1} \end{APACrefDOI}
\PrintBackRefs{\CurrentBib}

\bibitem [\protect \citeauthoryear {%
Epstein%
}{%
Epstein%
}{%
{\protect \APACyear {1969}}%
}]{%
Epstein}
\APACinsertmetastar {%
Epstein}%
\begin{APACrefauthors}%
Epstein, E\BPBI S.%
\end{APACrefauthors}%
\unskip\
\newblock
\APACrefYearMonthDay{1969}{}{}.
\newblock
{\BBOQ}\APACrefatitle {Stochastic dynamic prediction} {Stochastic dynamic prediction}.{\BBCQ}
\newblock
\APACjournalVolNumPages{Tellus}{}{}{739 - 759}.
\newblock
\begin{APACrefDOI} \doi{10.1111/j.2153-3490.1969.tb00483.x} \end{APACrefDOI}
\PrintBackRefs{\CurrentBib}

\bibitem [\protect \citeauthoryear {%
Fisher%
}{%
Fisher%
}{%
{\protect \APACyear {1915}}%
}]{%
fisherz}
\APACinsertmetastar {%
fisherz}%
\begin{APACrefauthors}%
Fisher, R\BPBI A.%
\end{APACrefauthors}%
\unskip\
\newblock
\APACrefYearMonthDay{1915}{}{}.
\newblock
{\BBOQ}\APACrefatitle {Frequency Distribution of the Values of the Correlation Coefficient in Samples from an Indefinitely Large Population} {Frequency distribution of the values of the correlation coefficient in samples from an indefinitely large population}.{\BBCQ}
\newblock
\APACjournalVolNumPages{Biometrika}{10}{4}{507--521}.
\PrintBackRefs{\CurrentBib}

\bibitem [\protect \citeauthoryear {%
Golaz%
\ \protect \BOthers {.}}{%
Golaz%
\ \protect \BOthers {.}}{%
{\protect \APACyear {2022}}%
}]{%
golaz}
\APACinsertmetastar {%
golaz}%
\begin{APACrefauthors}%
Golaz, J\BHBI C.%
, Van~Roekel, L\BPBI P.%
, Zheng, X.%
, Roberts, A\BPBI F.%
, Wolfe, J\BPBI D.%
, Lin, W.%
\BDBL {}Bader, D\BPBI C.%
\end{APACrefauthors}%
\unskip\
\newblock
\APACrefYearMonthDay{2022}{}{}.
\newblock
{\BBOQ}\APACrefatitle {The DOE E3SM Model Version 2: Overview of the Physical Model and Initial Model Evaluation} {The doe e3sm model version 2: Overview of the physical model and initial model evaluation}.{\BBCQ}
\newblock
\APACjournalVolNumPages{Journal of Advances in Modeling Earth Systems}{14}{12}{e2022MS003156}.
\newblock
\begin{APACrefDOI} \doi{https://doi.org/10.1029/2022MS003156} \end{APACrefDOI}
\PrintBackRefs{\CurrentBib}

\bibitem [\protect \citeauthoryear {%
Goswami%
, Khouider%
, Phani%
, Mukhopadhyay%
\BCBL {}\ \BBA {} Majda%
}{%
Goswami%
\ \protect \BOthers {.}}{%
{\protect \APACyear {2017}}%
}]{%
goswami}
\APACinsertmetastar {%
goswami}%
\begin{APACrefauthors}%
Goswami, B\BPBI B.%
, Khouider, B.%
, Phani, R.%
, Mukhopadhyay, P.%
\BCBL {}\ \BBA {} Majda, A\BPBI J.%
\end{APACrefauthors}%
\unskip\
\newblock
\APACrefYearMonthDay{2017}{}{}.
\newblock
{\BBOQ}\APACrefatitle {Implementation and calibration of a stochastic multicloud convective parameterization in the NCEP Climate Forecast System (CFSv2)} {Implementation and calibration of a stochastic multicloud convective parameterization in the ncep climate forecast system (cfsv2)}.{\BBCQ}
\newblock
\APACjournalVolNumPages{Journal of Advances in Modeling Earth Systems}{9}{3}{1721-1739}.
\newblock
\begin{APACrefDOI} \doi{https://doi.org/10.1002/2017MS001014} \end{APACrefDOI}
\PrintBackRefs{\CurrentBib}

\bibitem [\protect \citeauthoryear {%
Ham%
, Kim%
\BCBL {}\ \BBA {} Luo%
}{%
Ham%
\ \protect \BOthers {.}}{%
{\protect \APACyear {2019}}%
}]{%
Ham}
\APACinsertmetastar {%
Ham}%
\begin{APACrefauthors}%
Ham, Y.%
, Kim, J.%
\BCBL {}\ \BBA {} Luo, J.%
\end{APACrefauthors}%
\unskip\
\newblock
\APACrefYearMonthDay{2019}{}{}.
\newblock
{\BBOQ}\APACrefatitle {Deep learning for multi-year ENSO forecasts} {Deep learning for multi-year enso forecasts}.{\BBCQ}
\newblock
\APACjournalVolNumPages{Nature}{573}{}{568-572}.
\newblock
\begin{APACrefDOI} \doi{10.1038/s41586-019-1559-7} \end{APACrefDOI}
\PrintBackRefs{\CurrentBib}

\bibitem [\protect \citeauthoryear {%
Harris%
\ \protect \BOthers {.}}{%
Harris%
\ \protect \BOthers {.}}{%
{\protect \APACyear {2020}}%
}]{%
harris2020array}
\APACinsertmetastar {%
harris2020array}%
\begin{APACrefauthors}%
Harris, C\BPBI R.%
, Millman, K\BPBI J.%
, van~der Walt, S\BPBI J.%
, Gommers, R.%
, Virtanen, P.%
, Cournapeau, D.%
\BDBL {}Oliphant, T\BPBI E.%
\end{APACrefauthors}%
\unskip\
\newblock
\APACrefYearMonthDay{2020}{{\APACmonth{09}}}{}.
\newblock
{\BBOQ}\APACrefatitle {Array programming with {NumPy}} {Array programming with {NumPy}}.{\BBCQ}
\newblock
\APACjournalVolNumPages{Nature}{585}{7825}{357--362}.
\newblock
\begin{APACrefURL} \url{https://doi.org/10.1038/s41586-020-2649-2} \end{APACrefURL}
\newblock
\begin{APACrefDOI} \doi{10.1038/s41586-020-2649-2} \end{APACrefDOI}
\PrintBackRefs{\CurrentBib}

\bibitem [\protect \citeauthoryear {%
Hersbach%
\ \protect \BOthers {.}}{%
Hersbach%
\ \protect \BOthers {.}}{%
{\protect \APACyear {2020}}%
}]{%
Hersbach}
\APACinsertmetastar {%
Hersbach}%
\begin{APACrefauthors}%
Hersbach, H.%
, Bell, B.%
, Berrisford, P.%
, Hirahara, S.%
, Horanyi, A.%
, Mu\~noz Sabater, J.%
\BDBL {}Th\'epaut, J\BHBI N.%
\end{APACrefauthors}%
\unskip\
\newblock
\APACrefYearMonthDay{2020}{}{}.
\newblock
{\BBOQ}\APACrefatitle {The ERA5 global reanalysis} {The era5 global reanalysis}.{\BBCQ}
\newblock
\APACjournalVolNumPages{Quarterly Journal of the Royal Meteorological Society}{146}{730}{1999-2049}.
\newblock
\begin{APACrefDOI} \doi{https://doi.org/10.1002/qj.3803} \end{APACrefDOI}
\PrintBackRefs{\CurrentBib}

\bibitem [\protect \citeauthoryear {%
Hunter%
}{%
Hunter%
}{%
{\protect \APACyear {2007}}%
}]{%
Hunter:2007}
\APACinsertmetastar {%
Hunter:2007}%
\begin{APACrefauthors}%
Hunter, J\BPBI D.%
\end{APACrefauthors}%
\unskip\
\newblock
\APACrefYearMonthDay{2007}{}{}.
\newblock
{\BBOQ}\APACrefatitle {Matplotlib: A 2D graphics environment} {Matplotlib: A 2d graphics environment}.{\BBCQ}
\newblock
\APACjournalVolNumPages{Computing in Science \& Engineering}{9}{3}{90--95}.
\newblock
\begin{APACrefDOI} \doi{10.1109/MCSE.2007.55} \end{APACrefDOI}
\PrintBackRefs{\CurrentBib}

\bibitem [\protect \citeauthoryear {%
Jaeger%
}{%
Jaeger%
}{%
{\protect \APACyear {2001}}%
}]{%
Jaeger}
\APACinsertmetastar {%
Jaeger}%
\begin{APACrefauthors}%
Jaeger, H.%
\end{APACrefauthors}%
\unskip\
\newblock
\APACrefYearMonthDay{2001}{}{}.
\newblock
{\BBOQ}\APACrefatitle {The “echo state” approach to analysing and training recurrent neural networks} {The “echo state” approach to analysing and training recurrent neural networks}.{\BBCQ}
\newblock
\APACjournalVolNumPages{GMD Report, German National Research Center for Information Technology}{148}{}{}.
\PrintBackRefs{\CurrentBib}

\bibitem [\protect \citeauthoryear {%
Ji%
\ \BBA {} Leetmaa%
}{%
Ji%
\ \BBA {} Leetmaa%
}{%
{\protect \APACyear {1997}}%
}]{%
Ji}
\APACinsertmetastar {%
Ji}%
\begin{APACrefauthors}%
Ji, M.%
\BCBT {}\ \BBA {} Leetmaa, A.%
\end{APACrefauthors}%
\unskip\
\newblock
\APACrefYearMonthDay{1997}{}{}.
\newblock
{\BBOQ}\APACrefatitle {Impact of Data Assimilation on Ocean Initialization and El Niño Prediction} {Impact of data assimilation on ocean initialization and el niño prediction}.{\BBCQ}
\newblock
\APACjournalVolNumPages{Monthly Weather Review}{125}{5}{742-753}.
\newblock
\begin{APACrefDOI} \doi{https://doi.org/10.1175/1520-0493(1997)125<0742:IODAOO>2.0.CO;2} \end{APACrefDOI}
\PrintBackRefs{\CurrentBib}

\bibitem [\protect \citeauthoryear {%
Kirtman%
\ \protect \BOthers {.}}{%
Kirtman%
\ \protect \BOthers {.}}{%
{\protect \APACyear {2014}}%
}]{%
Kirtman}
\APACinsertmetastar {%
Kirtman}%
\begin{APACrefauthors}%
Kirtman, B\BPBI P.%
, Min, D.%
, Infanti, J\BPBI M.%
, Kinter, J\BPBI L.%
, Paolino, D\BPBI A.%
, Zhang, Q.%
\BDBL {}Wood, E\BPBI F.%
\end{APACrefauthors}%
\unskip\
\newblock
\APACrefYearMonthDay{2014}{}{}.
\newblock
{\BBOQ}\APACrefatitle {The North American Multimodel Ensemble: Phase-1 Seasonal-to-Interannual Prediction; Phase-2 toward Developing Intraseasonal Prediction} {The north american multimodel ensemble: Phase-1 seasonal-to-interannual prediction; phase-2 toward developing intraseasonal prediction}.{\BBCQ}
\newblock
\APACjournalVolNumPages{Bulletin of the American Meteorological Society}{95}{4}{585 - 601}.
\newblock
\begin{APACrefDOI} \doi{10.1175/BAMS-D-12-00050.1} \end{APACrefDOI}
\PrintBackRefs{\CurrentBib}

\bibitem [\protect \citeauthoryear {%
Kochkov%
\ \protect \BOthers {.}}{%
Kochkov%
\ \protect \BOthers {.}}{%
{\protect \APACyear {2023}}%
}]{%
kochkov2023neural}
\APACinsertmetastar {%
kochkov2023neural}%
\begin{APACrefauthors}%
Kochkov, D.%
, Yuval, J.%
, Langmore, I.%
, Norgaard, P.%
, Smith, J.%
, Mooers, G.%
\BDBL {}Hoyer, S.%
\end{APACrefauthors}%
\unskip\
\newblock
\APACrefYearMonthDay{2023}{}{}.
\newblock
\APACrefbtitle {Neural General Circulation Models.} {Neural general circulation models.}
\newblock
\begin{APACrefDOI} \doi{https://doi.org/10.48550/arXiv.2311.07222} \end{APACrefDOI}
\PrintBackRefs{\CurrentBib}

\bibitem [\protect \citeauthoryear {%
Kucharski%
, Molteni%
\BCBL {}\ \BBA {} Bracco%
}{%
Kucharski%
\ \protect \BOthers {.}}{%
{\protect \APACyear {2006}}%
}]{%
Kucharski}
\APACinsertmetastar {%
Kucharski}%
\begin{APACrefauthors}%
Kucharski, F.%
, Molteni, F.%
\BCBL {}\ \BBA {} Bracco, A.%
\end{APACrefauthors}%
\unskip\
\newblock
\APACrefYearMonthDay{2006}{}{}.
\newblock
{\BBOQ}\APACrefatitle {Decadal interactions between the western tropical Pacific and the North Atlantic Oscillation} {Decadal interactions between the western tropical pacific and the north atlantic oscillation}.{\BBCQ}
\newblock
\APACjournalVolNumPages{Climate Dynamics}{26}{}{79-91}.
\newblock
\begin{APACrefDOI} \doi{https://doi.org/10.1007/s00382-005-0085-5} \end{APACrefDOI}
\PrintBackRefs{\CurrentBib}

\bibitem [\protect \citeauthoryear {%
Kwa%
\ \protect \BOthers {.}}{%
Kwa%
\ \protect \BOthers {.}}{%
{\protect \APACyear {2023}}%
}]{%
Kwa2023}
\APACinsertmetastar {%
Kwa2023}%
\begin{APACrefauthors}%
Kwa, A.%
, Clark, S\BPBI K.%
, Henn, B.%
, Brenowitz, N\BPBI D.%
, McGibbon, J.%
, Watt-Meyer, O.%
\BDBL {}Bretherton, C\BPBI S.%
\end{APACrefauthors}%
\unskip\
\newblock
\APACrefYearMonthDay{2023}{}{}.
\newblock
{\BBOQ}\APACrefatitle {Machine-Learned Climate Model Corrections From a Global Storm-Resolving Model: Performance Across the Annual Cycle} {Machine-learned climate model corrections from a global storm-resolving model: Performance across the annual cycle}.{\BBCQ}
\newblock
\APACjournalVolNumPages{Journal of Advances in Modeling Earth Systems}{15}{5}{e2022MS003400}.
\newblock
\begin{APACrefDOI} \doi{https://doi.org/10.1029/2022MS003400} \end{APACrefDOI}
\PrintBackRefs{\CurrentBib}

\bibitem [\protect \citeauthoryear {%
Lai%
, Herzog%
\BCBL {}\ \BBA {} Graf%
}{%
Lai%
\ \protect \BOthers {.}}{%
{\protect \APACyear {2018}}%
}]{%
Lai}
\APACinsertmetastar {%
Lai}%
\begin{APACrefauthors}%
Lai, A\BPBI W\BHBI C.%
, Herzog, M.%
\BCBL {}\ \BBA {} Graf, H\BHBI F.%
\end{APACrefauthors}%
\unskip\
\newblock
\APACrefYearMonthDay{2018}{}{}.
\newblock
{\BBOQ}\APACrefatitle {ENSO Forecasts near the Spring Predictability Barrier and Possible Reasons for the Recently Reduced Predictability} {Enso forecasts near the spring predictability barrier and possible reasons for the recently reduced predictability}.{\BBCQ}
\newblock
\APACjournalVolNumPages{Journal of Climate}{31}{2}{815 - 838}.
\newblock
\begin{APACrefDOI} \doi{10.1175/JCLI-D-17-0180.1} \end{APACrefDOI}
\PrintBackRefs{\CurrentBib}

\bibitem [\protect \citeauthoryear {%
Lam%
\ \protect \BOthers {.}}{%
Lam%
\ \protect \BOthers {.}}{%
{\protect \APACyear {2023}}%
}]{%
Lam2023}
\APACinsertmetastar {%
Lam2023}%
\begin{APACrefauthors}%
Lam, R.%
, Sanchez-Gonzalez, A.%
, Willson, M.%
, Wirnsberger, P.%
, Fortunato, M.%
, Alet, F.%
\BDBL {}Battaglia, P.%
\end{APACrefauthors}%
\unskip\
\newblock
\APACrefYearMonthDay{2023}{}{}.
\newblock
{\BBOQ}\APACrefatitle {Learning skillful medium-range global weather forecasting} {Learning skillful medium-range global weather forecasting}.{\BBCQ}
\newblock
\APACjournalVolNumPages{Science}{382}{}{1416--1421}.
\newblock
\begin{APACrefDOI} \doi{10.1126/science.adi2336} \end{APACrefDOI}
\PrintBackRefs{\CurrentBib}

\bibitem [\protect \citeauthoryear {%
Lavers%
, Simmons%
, Vamborg%
\BCBL {}\ \BBA {} Rodwell%
}{%
Lavers%
\ \protect \BOthers {.}}{%
{\protect \APACyear {2022}}%
}]{%
Lavers}
\APACinsertmetastar {%
Lavers}%
\begin{APACrefauthors}%
Lavers, D\BPBI A.%
, Simmons, A.%
, Vamborg, F.%
\BCBL {}\ \BBA {} Rodwell, M\BPBI J.%
\end{APACrefauthors}%
\unskip\
\newblock
\APACrefYearMonthDay{2022}{}{}.
\newblock
{\BBOQ}\APACrefatitle {An evaluation of ERA5 precipitation for climate monitoring} {An evaluation of era5 precipitation for climate monitoring}.{\BBCQ}
\newblock
\APACjournalVolNumPages{Quarterly Journal of the Royal Meteorological Society}{148}{748}{3152-3165}.
\newblock
\begin{APACrefDOI} \doi{https://doi.org/10.1002/qj.4351} \end{APACrefDOI}
\PrintBackRefs{\CurrentBib}

\bibitem [\protect \citeauthoryear {%
Lee%
\ \protect \BOthers {.}}{%
Lee%
\ \protect \BOthers {.}}{%
{\protect \APACyear {2022}}%
}]{%
lee}
\APACinsertmetastar {%
lee}%
\begin{APACrefauthors}%
Lee, S.%
, L'Heureux, M.%
, Wittenberg, A\BPBI T.%
, Seager, R.%
, O'Gorman, P\BPBI A.%
\BCBL {}\ \BBA {} Johnson, N\BPBI C.%
\end{APACrefauthors}%
\unskip\
\newblock
\APACrefYearMonthDay{2022}{}{}.
\newblock
{\BBOQ}\APACrefatitle {On the future zonal contrasts of equatorial Pacific climate: Perspectives from Observations, Simulations, and Theories} {On the future zonal contrasts of equatorial pacific climate: Perspectives from observations, simulations, and theories}.{\BBCQ}
\newblock
\APACjournalVolNumPages{npj Clim Atmos Sci}{5}{82}{}.
\newblock
\begin{APACrefDOI} \doi{10.1038/s41612-022-00301-2} \end{APACrefDOI}
\PrintBackRefs{\CurrentBib}

\bibitem [\protect \citeauthoryear {%
Leith%
}{%
Leith%
}{%
{\protect \APACyear {1974}}%
}]{%
Leith}
\APACinsertmetastar {%
Leith}%
\begin{APACrefauthors}%
Leith, C\BPBI E.%
\end{APACrefauthors}%
\unskip\
\newblock
\APACrefYearMonthDay{1974}{}{}.
\newblock
{\BBOQ}\APACrefatitle {Theoretical skill of {M}onte {C}arlo forecasts} {Theoretical skill of {M}onte {C}arlo forecasts}.{\BBCQ}
\newblock
\APACjournalVolNumPages{Monthly Weather Review}{102}{6}{409 - 418}.
\newblock
\begin{APACrefDOI} \doi{10.1175/1520-0493(1974)102%3C0409:TSOMCF%3E2.0.CO;2} \end{APACrefDOI}
\PrintBackRefs{\CurrentBib}

\bibitem [\protect \citeauthoryear {%
Li%
\ \BBA {} Robertson%
}{%
Li%
\ \BBA {} Robertson%
}{%
{\protect \APACyear {2015}}%
}]{%
Li}
\APACinsertmetastar {%
Li}%
\begin{APACrefauthors}%
Li, S.%
\BCBT {}\ \BBA {} Robertson, A\BPBI W.%
\end{APACrefauthors}%
\unskip\
\newblock
\APACrefYearMonthDay{2015}{}{}.
\newblock
{\BBOQ}\APACrefatitle {Evaluation of Submonthly Precipitation Forecast Skill from Global Ensemble Prediction Systems} {Evaluation of submonthly precipitation forecast skill from global ensemble prediction systems}.{\BBCQ}
\newblock
\APACjournalVolNumPages{Monthly Weather Review}{143}{7}{2871 - 2889}.
\newblock
\begin{APACrefDOI} \doi{10.1175/MWR-D-14-00277.1} \end{APACrefDOI}
\PrintBackRefs{\CurrentBib}

\bibitem [\protect \citeauthoryear {%
Lin%
\ \BBA {} Qian%
}{%
Lin%
\ \BBA {} Qian%
}{%
{\protect \APACyear {2019}}%
}]{%
Lin}
\APACinsertmetastar {%
Lin}%
\begin{APACrefauthors}%
Lin, J.%
\BCBT {}\ \BBA {} Qian, T.%
\end{APACrefauthors}%
\unskip\
\newblock
\APACrefYearMonthDay{2019}{}{}.
\newblock
{\BBOQ}\APACrefatitle {A New Picture of the Global Impacts of El Nino-Southern} {A new picture of the global impacts of el nino-southern}.{\BBCQ}
\newblock
\APACjournalVolNumPages{Scientific Reports}{}{9}{17543}.
\newblock
\begin{APACrefDOI} \doi{10.1038/s41598-019-54090-5} \end{APACrefDOI}
\PrintBackRefs{\CurrentBib}

\bibitem [\protect \citeauthoryear {%
Lindzen%
}{%
Lindzen%
}{%
{\protect \APACyear {1967}}%
}]{%
lindzen}
\APACinsertmetastar {%
lindzen}%
\begin{APACrefauthors}%
Lindzen, R\BPBI D.%
\end{APACrefauthors}%
\unskip\
\newblock
\APACrefYearMonthDay{1967}{}{}.
\newblock
{\BBOQ}\APACrefatitle {PLANETARY WAVES ON BETA-PLANES} {Planetary waves on beta-planes}.{\BBCQ}
\newblock
\APACjournalVolNumPages{Monthly Weather Review}{95}{7}{441 - 451}.
\newblock
\begin{APACrefDOI} \doi{10.1175/1520-0493(1967)095<0441:PWOBP>2.3.CO;2} \end{APACrefDOI}
\PrintBackRefs{\CurrentBib}

\bibitem [\protect \citeauthoryear {%
Liu%
, Cai%
, Lin%
, Li%
\BCBL {}\ \BBA {} Zhang%
}{%
Liu%
\ \protect \BOthers {.}}{%
{\protect \APACyear {2023}}%
}]{%
Liu}
\APACinsertmetastar {%
Liu}%
\begin{APACrefauthors}%
Liu, Y.%
, Cai, W.%
, Lin, X.%
, Li, Z.%
\BCBL {}\ \BBA {} Zhang, Y.%
\end{APACrefauthors}%
\unskip\
\newblock
\APACrefYearMonthDay{2023}{}{}.
\newblock
{\BBOQ}\APACrefatitle {Nonlinear El Nino impacts on the global economy under climate change} {Nonlinear el nino impacts on the global economy under climate change}.{\BBCQ}
\newblock
\APACjournalVolNumPages{Nature Communications}{}{14}{5887}.
\newblock
\begin{APACrefDOI} \doi{10.1038/s41467-023-41551-9} \end{APACrefDOI}
\PrintBackRefs{\CurrentBib}

\bibitem [\protect \citeauthoryear {%
Losada%
\ \protect \BOthers {.}}{%
Losada%
\ \protect \BOthers {.}}{%
{\protect \APACyear {2012}}%
}]{%
Losada}
\APACinsertmetastar {%
Losada}%
\begin{APACrefauthors}%
Losada, T.%
, Rodriguez-Fonseca, B.%
, Mohino, E.%
, Bader, J.%
, Janicot, S.%
\BCBL {}\ \BBA {} Mechoso, C\BPBI R.%
\end{APACrefauthors}%
\unskip\
\newblock
\APACrefYearMonthDay{2012}{}{}.
\newblock
{\BBOQ}\APACrefatitle {Tropical SST and Sahel rainfall: A non-stationary relationship} {Tropical sst and sahel rainfall: A non-stationary relationship}.{\BBCQ}
\newblock
\APACjournalVolNumPages{Geophysical Research Letters}{39}{12}{}.
\newblock
\begin{APACrefDOI} \doi{https://doi.org/10.1029/2012GL052423} \end{APACrefDOI}
\PrintBackRefs{\CurrentBib}

\bibitem [\protect \citeauthoryear {%
Luko{\v{s}}evi{\v{c}}ius%
}{%
Luko{\v{s}}evi{\v{c}}ius%
}{%
{\protect \APACyear {2012}}%
}]{%
Luk}
\APACinsertmetastar {%
Luk}%
\begin{APACrefauthors}%
Luko{\v{s}}evi{\v{c}}ius, M.%
\end{APACrefauthors}%
\unskip\
\newblock
\APACrefYearMonthDay{2012}{}{}.
\newblock
{\BBOQ}\APACrefatitle {A Practical Guide to Applying Echo State Networks} {A practical guide to applying echo state networks}.{\BBCQ}
\newblock
\BIn{} G.~Montavon, G\BPBI B.~Orr\BCBL {}\ \BBA {} K\BHBI R.~M{\"u}ller\ (\BEDS), \APACrefbtitle {Neural Networks: Tricks of the Trade: Second Edition} {Neural networks: Tricks of the trade: Second edition}\ (\BPGS\ 659--686).
\newblock
\APACaddressPublisher{Berlin, Heidelberg}{Springer Berlin Heidelberg}.
\newblock
\begin{APACrefDOI} \doi{10.1007/978-3-642-35289-8_36} \end{APACrefDOI}
\PrintBackRefs{\CurrentBib}

\bibitem [\protect \citeauthoryear {%
Lukoševičius%
\ \BBA {} Jaeger%
}{%
Lukoševičius%
\ \BBA {} Jaeger%
}{%
{\protect \APACyear {2009}}%
}]{%
Luk+Jaeger}
\APACinsertmetastar {%
Luk+Jaeger}%
\begin{APACrefauthors}%
Lukoševičius, M.%
\BCBT {}\ \BBA {} Jaeger, H.%
\end{APACrefauthors}%
\unskip\
\newblock
\APACrefYearMonthDay{2009}{}{}.
\newblock
{\BBOQ}\APACrefatitle {Reservoir computing approaches to recurrent neural network training} {Reservoir computing approaches to recurrent neural network training}.{\BBCQ}
\newblock
\APACjournalVolNumPages{Computer Science Review}{3}{3}{127-149}.
\newblock
\begin{APACrefDOI} \doi{https://doi.org/10.1016/j.cosrev.2009.03.005} \end{APACrefDOI}
\PrintBackRefs{\CurrentBib}

\bibitem [\protect \citeauthoryear {%
Mason%
\ \BBA {} Goddard%
}{%
Mason%
\ \BBA {} Goddard%
}{%
{\protect \APACyear {2001}}%
}]{%
Mason}
\APACinsertmetastar {%
Mason}%
\begin{APACrefauthors}%
Mason, S\BPBI J.%
\BCBT {}\ \BBA {} Goddard, L.%
\end{APACrefauthors}%
\unskip\
\newblock
\APACrefYearMonthDay{2001}{}{}.
\newblock
{\BBOQ}\APACrefatitle {Probabilistic Precipitation Anomalies Associated with EN SO} {Probabilistic precipitation anomalies associated with en so}.{\BBCQ}
\newblock
\APACjournalVolNumPages{Bulletin of the American Meteorological Society}{82}{4}{619 - 638}.
\newblock
\begin{APACrefDOI} \doi{10.1175/1520-0477(2001)082<0619:PPAAWE>2.3.CO;2} \end{APACrefDOI}
\PrintBackRefs{\CurrentBib}

\bibitem [\protect \citeauthoryear {%
Matsuno%
}{%
Matsuno%
}{%
{\protect \APACyear {1966}}%
}]{%
matsuno}
\APACinsertmetastar {%
matsuno}%
\begin{APACrefauthors}%
Matsuno, T.%
\end{APACrefauthors}%
\unskip\
\newblock
\APACrefYearMonthDay{1966}{}{}.
\newblock
{\BBOQ}\APACrefatitle {Quasi-Geostrophic Motions in the Equatorial Area} {Quasi-geostrophic motions in the equatorial area}.{\BBCQ}
\newblock
\APACjournalVolNumPages{Journal of the Meteorological Society of Japan. Ser. II}{44}{1}{25-43}.
\newblock
\begin{APACrefDOI} \doi{10.2151/jmsj1965.44.1_25} \end{APACrefDOI}
\PrintBackRefs{\CurrentBib}

\bibitem [\protect \citeauthoryear {%
McPhaden%
}{%
McPhaden%
}{%
{\protect \APACyear {2003}}%
}]{%
McPhaden}
\APACinsertmetastar {%
McPhaden}%
\begin{APACrefauthors}%
McPhaden, M\BPBI J.%
\end{APACrefauthors}%
\unskip\
\newblock
\APACrefYearMonthDay{2003}{}{}.
\newblock
{\BBOQ}\APACrefatitle {Tropical Pacific Ocean heat content variations and ENSO persistence barriers} {Tropical pacific ocean heat content variations and enso persistence barriers}.{\BBCQ}
\newblock
\APACjournalVolNumPages{Geophysical Research Letters}{30}{9}{}.
\newblock
\begin{APACrefDOI} \doi{https://doi.org/10.1029/2003GL016872} \end{APACrefDOI}
\PrintBackRefs{\CurrentBib}

\bibitem [\protect \citeauthoryear {%
McPhaden%
, Zebiak%
\BCBL {}\ \BBA {} Glantz%
}{%
McPhaden%
\ \protect \BOthers {.}}{%
{\protect \APACyear {2006}}%
}]{%
McPhaden1}
\APACinsertmetastar {%
McPhaden1}%
\begin{APACrefauthors}%
McPhaden, M\BPBI J.%
, Zebiak, S\BPBI E.%
\BCBL {}\ \BBA {} Glantz, M\BPBI H.%
\end{APACrefauthors}%
\unskip\
\newblock
\APACrefYearMonthDay{2006}{}{}.
\newblock
{\BBOQ}\APACrefatitle {ENSO as an Integrating Concept in Earth Science} {Enso as an integrating concept in earth science}.{\BBCQ}
\newblock
\APACjournalVolNumPages{Science}{314}{5806}{1740-1745}.
\newblock
\begin{APACrefDOI} \doi{10.1126/science.1132588} \end{APACrefDOI}
\PrintBackRefs{\CurrentBib}

\bibitem [\protect \citeauthoryear {%
{Met Office}%
}{%
{Met Office}%
}{%
{\protect \APACyear {2010 - 2015}}%
}]{%
Cartopy}
\APACinsertmetastar {%
Cartopy}%
\begin{APACrefauthors}%
{Met Office}.%
\end{APACrefauthors}%
\unskip\
\newblock
\APACrefYearMonthDay{2010 - 2015}{}{}.
\newblock
{\BBOQ}\APACrefatitle {Cartopy: a cartographic python library with a Matplotlib interface} {Cartopy: a cartographic python library with a matplotlib interface}{\BBCQ}\ [\bibcomputersoftwaremanual].
\newblock
\APACaddressPublisher{Exeter, Devon}{}.
\newblock
\begin{APACrefURL} \url{https://scitools.org.uk/cartopy} \end{APACrefURL}
\PrintBackRefs{\CurrentBib}

\bibitem [\protect \citeauthoryear {%
Molod%
\ \protect \BOthers {.}}{%
Molod%
\ \protect \BOthers {.}}{%
{\protect \APACyear {2020}}%
}]{%
molod}
\APACinsertmetastar {%
molod}%
\begin{APACrefauthors}%
Molod, A.%
, Hackert, E.%
, Vikhliaev, Y.%
, Zhao, B.%
, Barahona, D.%
, Vernieres, G.%
\BDBL {}Pawson, S.%
\end{APACrefauthors}%
\unskip\
\newblock
\APACrefYearMonthDay{2020}{}{}.
\newblock
{\BBOQ}\APACrefatitle {GEOS-S2S Version 2: The GMAO High-Resolution Coupled Model and Assimilation System for Seasonal Prediction} {Geos-s2s version 2: The gmao high-resolution coupled model and assimilation system for seasonal prediction}.{\BBCQ}
\newblock
\APACjournalVolNumPages{Journal of Geophysical Research: Atmospheres}{125}{5}{e2019JD031767}.
\newblock
\begin{APACrefDOI} \doi{https://doi.org/10.1029/2019JD031767} \end{APACrefDOI}
\PrintBackRefs{\CurrentBib}

\bibitem [\protect \citeauthoryear {%
Molteni%
}{%
Molteni%
}{%
{\protect \APACyear {2003}}%
}]{%
Molteni}
\APACinsertmetastar {%
Molteni}%
\begin{APACrefauthors}%
Molteni, F.%
\end{APACrefauthors}%
\unskip\
\newblock
\APACrefYearMonthDay{2003}{}{}.
\newblock
{\BBOQ}\APACrefatitle {Atmospheric simulations using a GCM with simplified physical parametrizations. I: model climatology and variability in multi-decadal experiments} {Atmospheric simulations using a gcm with simplified physical parametrizations. i: model climatology and variability in multi-decadal experiments}.{\BBCQ}
\newblock
\APACjournalVolNumPages{Climate Dynamics}{20}{}{175-191}.
\newblock
\begin{APACrefDOI} \doi{https://doi.org/10.1007/s00382-002-0268-2} \end{APACrefDOI}
\PrintBackRefs{\CurrentBib}

\bibitem [\protect \citeauthoryear {%
Monhart%
\ \protect \BOthers {.}}{%
Monhart%
\ \protect \BOthers {.}}{%
{\protect \APACyear {2018}}%
}]{%
Monhart}
\APACinsertmetastar {%
Monhart}%
\begin{APACrefauthors}%
Monhart, S.%
, Spirig, C.%
, Bhend, J.%
, Bogner, K.%
, Schär, C.%
\BCBL {}\ \BBA {} Liniger, M\BPBI A.%
\end{APACrefauthors}%
\unskip\
\newblock
\APACrefYearMonthDay{2018}{}{}.
\newblock
{\BBOQ}\APACrefatitle {Skill of Subseasonal Forecasts in Europe: Effect of Bias Correction and Downscaling Using Surface Observations} {Skill of subseasonal forecasts in europe: Effect of bias correction and downscaling using surface observations}.{\BBCQ}
\newblock
\APACjournalVolNumPages{Journal of Geophysical Research: Atmospheres}{123}{15}{7999-8016}.
\newblock
\begin{APACrefDOI} \doi{https://doi.org/10.1029/2017JD027923} \end{APACrefDOI}
\PrintBackRefs{\CurrentBib}

\bibitem [\protect \citeauthoryear {%
Mouatadid%
\ \protect \BOthers {.}}{%
Mouatadid%
\ \protect \BOthers {.}}{%
{\protect \APACyear {2023}}%
}]{%
Mouatadid}
\APACinsertmetastar {%
Mouatadid}%
\begin{APACrefauthors}%
Mouatadid, S.%
, Orenstein, P.%
, Flaspohler, G.%
, Cohen, J.%
, Oprescu, M.%
, Fraenkel, E.%
\BCBL {}\ \BBA {} Mackey, L.%
\end{APACrefauthors}%
\unskip\
\newblock
\APACrefYearMonthDay{2023}{}{}.
\newblock
{\BBOQ}\APACrefatitle {Adaptive bias correction for improved subseasonal forecasting.} {Adaptive bias correction for improved subseasonal forecasting.}{\BBCQ}
\newblock
\APACjournalVolNumPages{Nature Communications}{14}{}{3482}.
\newblock
\begin{APACrefDOI} \doi{https://doi.org/10.1038/s41467-023-38874-y} \end{APACrefDOI}
\PrintBackRefs{\CurrentBib}

\bibitem [\protect \citeauthoryear {%
{National Academies of Sciences and Engineering and Medicine}%
}{%
{National Academies of Sciences and Engineering and Medicine}%
}{%
{\protect \APACyear {2016}}%
}]{%
NAP21873}
\APACinsertmetastar {%
NAP21873}%
\begin{APACrefauthors}%
{National Academies of Sciences and Engineering and Medicine}.%
\end{APACrefauthors}%
\unskip\
\newblock
\APACrefYear{2016}.
\newblock
\APACrefbtitle {Next Generation Earth System Prediction: Strategies for Subseasonal to Seasonal Forecasts} {Next generation earth system prediction: Strategies for subseasonal to seasonal forecasts}.
\newblock
\APACaddressPublisher{Washington, DC}{The National Academies Press}.
\newblock
\begin{APACrefDOI} \doi{10.17226/21873} \end{APACrefDOI}
\PrintBackRefs{\CurrentBib}

\bibitem [\protect \citeauthoryear {%
Nguyen%
\ \protect \BOthers {.}}{%
Nguyen%
\ \protect \BOthers {.}}{%
{\protect \APACyear {2023}}%
}]{%
nguyen2023scaling}
\APACinsertmetastar {%
nguyen2023scaling}%
\begin{APACrefauthors}%
Nguyen, T.%
, Shah, R.%
, Bansal, H.%
, Arcomano, T.%
, Madireddy, S.%
, Maulik, R.%
\BDBL {}Grover, A.%
\end{APACrefauthors}%
\unskip\
\newblock
\APACrefYearMonthDay{2023}{}{}.
\newblock
\APACrefbtitle {Scaling transformer neural networks for skillful and reliable medium-range weather forecasting.} {Scaling transformer neural networks for skillful and reliable medium-range weather forecasting.}
\newblock
\begin{APACrefDOI} \doi{https://doi.org/10.48550/arXiv.2312.03876} \end{APACrefDOI}
\PrintBackRefs{\CurrentBib}

\bibitem [\protect \citeauthoryear {%
Owens%
\ \BBA {} Hewson%
}{%
Owens%
\ \BBA {} Hewson%
}{%
{\protect \APACyear {2018}}%
}]{%
Owens}
\APACinsertmetastar {%
Owens}%
\begin{APACrefauthors}%
Owens, R\BPBI G.%
\BCBT {}\ \BBA {} Hewson, T\BPBI D.%
\end{APACrefauthors}%
\unskip\
\newblock
\APACrefYearMonthDay{2018}{}{}.
\newblock
\APACrefbtitle {ECMWF Forecast User Guide. Reading: ECMWF.} {Ecmwf forecast user guide. reading: Ecmwf.}
\newblock
\begin{APACrefDOI} \doi{10.21957/m1cs7h} \end{APACrefDOI}
\PrintBackRefs{\CurrentBib}

\bibitem [\protect \citeauthoryear {%
Patel%
}{%
Patel%
}{%
{\protect \APACyear {2024}}%
{\protect \APACexlab {{\protect \BCnt {1}}}}}]{%
dhruvitdata}
\APACinsertmetastar {%
dhruvitdata}%
\begin{APACrefauthors}%
Patel, D.%
\end{APACrefauthors}%
\unskip\
\newblock
\APACrefYearMonthDay{2024{\protect \BCnt {1}}}{}{}.
\newblock
\APACrefbtitle {\textnormal{Hybrid Model Sample Data}.} {\textnormal{Hybrid Model Sample Data}.}
\newblock
\APAChowpublished {[Data set]}.
\newblock
\APACaddressPublisher{}{Zenodo}.
\newblock
\begin{APACrefURL} \url{https://doi.org/10.5281/zenodo.14043079} \end{APACrefURL}
\PrintBackRefs{\CurrentBib}

\bibitem [\protect \citeauthoryear {%
Patel%
}{%
Patel%
}{%
{\protect \APACyear {2024}}%
{\protect \APACexlab {{\protect \BCnt {2}}}}}]{%
dhruvitweights}
\APACinsertmetastar {%
dhruvitweights}%
\begin{APACrefauthors}%
Patel, D.%
\end{APACrefauthors}%
\unskip\
\newblock
\APACrefYearMonthDay{2024{\protect \BCnt {2}}}{}{}.
\newblock
\APACrefbtitle {\textnormal{Hybrid Model Training Weights}.} {\textnormal{Hybrid Model Training Weights}.}
\newblock
\APAChowpublished {[Data set]}.
\newblock
\APACaddressPublisher{}{Zenodo}.
\newblock
\begin{APACrefURL} \url{https://doi.org/10.5281/zenodo.11390866} \end{APACrefURL}
\PrintBackRefs{\CurrentBib}

\bibitem [\protect \citeauthoryear {%
Patel%
\ \BBA {} Arcomano%
}{%
Patel%
\ \BBA {} Arcomano%
}{%
{\protect \APACyear {2024}}%
}]{%
dhruvitcode}
\APACinsertmetastar {%
dhruvitcode}%
\begin{APACrefauthors}%
Patel, D.%
\BCBT {}\ \BBA {} Arcomano, T.%
\end{APACrefauthors}%
\unskip\
\newblock
\APACrefYearMonthDay{2024}{}{}.
\newblock
\APACrefbtitle {\textnormal{Hybrid-Weather-Beyond-Medium-Range}.} {\textnormal{Hybrid-Weather-Beyond-Medium-Range}.}
\newblock
\APAChowpublished {[Software]}.
\newblock
\APACaddressPublisher{}{Zenodo}.
\newblock
\begin{APACrefURL} \url{https://doi.org/10.5281/zenodo.14038114} \end{APACrefURL}
\PrintBackRefs{\CurrentBib}

\bibitem [\protect \citeauthoryear {%
Patel%
\ \BBA {} Ott%
}{%
Patel%
\ \BBA {} Ott%
}{%
{\protect \APACyear {2023}}%
}]{%
patel}
\APACinsertmetastar {%
patel}%
\begin{APACrefauthors}%
Patel, D.%
\BCBT {}\ \BBA {} Ott, E.%
\end{APACrefauthors}%
\unskip\
\newblock
\APACrefYearMonthDay{2023}{02}{}.
\newblock
{\BBOQ}\APACrefatitle {{Using machine learning to anticipate tipping points and extrapolate to post-tipping dynamics of non-stationary dynamical systems}} {{Using machine learning to anticipate tipping points and extrapolate to post-tipping dynamics of non-stationary dynamical systems}}.{\BBCQ}
\newblock
\APACjournalVolNumPages{Chaos: An Interdisciplinary Journal of Nonlinear Science}{33}{2}{023143}.
\newblock
\begin{APACrefDOI} \doi{10.1063/5.0131787} \end{APACrefDOI}
\PrintBackRefs{\CurrentBib}

\bibitem [\protect \citeauthoryear {%
Pathak%
, Hunt%
, Girvan%
, Lu%
\BCBL {}\ \BBA {} Ott%
}{%
Pathak%
, Hunt%
\BCBL {}\ \protect \BOthers {.}}{%
{\protect \APACyear {2018}}%
}]{%
pathakPRL}
\APACinsertmetastar {%
pathakPRL}%
\begin{APACrefauthors}%
Pathak, J.%
, Hunt, B.%
, Girvan, M.%
, Lu, Z.%
\BCBL {}\ \BBA {} Ott, E.%
\end{APACrefauthors}%
\unskip\
\newblock
\APACrefYearMonthDay{2018}{Jan}{}.
\newblock
{\BBOQ}\APACrefatitle {Model-Free Prediction of Large Spatiotemporally Chaotic Systems from Data: A Reservoir Computing Approach} {Model-free prediction of large spatiotemporally chaotic systems from data: A reservoir computing approach}.{\BBCQ}
\newblock
\APACjournalVolNumPages{Phys. Rev. Lett.}{120}{}{024102}.
\newblock
\begin{APACrefDOI} \doi{10.1103/PhysRevLett.120.024102} \end{APACrefDOI}
\PrintBackRefs{\CurrentBib}

\bibitem [\protect \citeauthoryear {%
Pathak%
\ \protect \BOthers {.}}{%
Pathak%
\ \protect \BOthers {.}}{%
{\protect \APACyear {2022}}%
}]{%
Pathak2022}
\APACinsertmetastar {%
Pathak2022}%
\begin{APACrefauthors}%
Pathak, J.%
, Subramanian, S.%
, Harrington, P.%
, Raja, S.%
, Chattopadhyay, A.%
, Mardani, M.%
\BDBL {}Anandkumar, A.%
\end{APACrefauthors}%
\unskip\
\newblock
\APACrefYearMonthDay{2022}{}{}.
\newblock
{\BBOQ}\APACrefatitle {FourCastNet: A Global Data-driven High-resolution Weather Model using Adaptive Fourier Neural Operators} {Fourcastnet: A global data-driven high-resolution weather model using adaptive fourier neural operators}.{\BBCQ}
\newblock
\APACjournalVolNumPages{arXiv}{}{}{}.
\newblock
\begin{APACrefDOI} \doi{https://doi.org/10.48550/arXiv.2202.11214} \end{APACrefDOI}
\PrintBackRefs{\CurrentBib}

\bibitem [\protect \citeauthoryear {%
Pathak%
, Wikner%
\BCBL {}\ \protect \BOthers {.}}{%
Pathak%
, Wikner%
\BCBL {}\ \protect \BOthers {.}}{%
{\protect \APACyear {2018}}%
}]{%
Pathak+Wikner}
\APACinsertmetastar {%
Pathak+Wikner}%
\begin{APACrefauthors}%
Pathak, J.%
, Wikner, A.%
, Fussell, R.%
, Chandra, S.%
, Hunt, B\BPBI R.%
, Girvan, M.%
\BCBL {}\ \BBA {} Ott, E.%
\end{APACrefauthors}%
\unskip\
\newblock
\APACrefYearMonthDay{2018}{}{}.
\newblock
{\BBOQ}\APACrefatitle {{Hybrid forecasting of chaotic processes: Using machine learning in conjunction with a knowledge-based model}} {{Hybrid forecasting of chaotic processes: Using machine learning in conjunction with a knowledge-based model}}.{\BBCQ}
\newblock
\APACjournalVolNumPages{Chaos: An Interdisciplinary Journal of Nonlinear Science}{28}{4}{041101}.
\newblock
\begin{APACrefDOI} \doi{10.1063/1.5028373} \end{APACrefDOI}
\PrintBackRefs{\CurrentBib}

\bibitem [\protect \citeauthoryear {%
Rasp%
, Pritchard%
\BCBL {}\ \BBA {} Gentine%
}{%
Rasp%
\ \protect \BOthers {.}}{%
{\protect \APACyear {2018}}%
}]{%
Rasp}
\APACinsertmetastar {%
Rasp}%
\begin{APACrefauthors}%
Rasp, S.%
, Pritchard, M\BPBI S.%
\BCBL {}\ \BBA {} Gentine, P.%
\end{APACrefauthors}%
\unskip\
\newblock
\APACrefYearMonthDay{2018}{}{}.
\newblock
{\BBOQ}\APACrefatitle {Deep learning to represent subgrid processes in climate models} {Deep learning to represent subgrid processes in climate models}.{\BBCQ}
\newblock
\APACjournalVolNumPages{Proceedings of the National Academy of Sciences}{115}{39}{9684-9689}.
\newblock
\begin{APACrefDOI} \doi{10.1073/pnas.1810286115} \end{APACrefDOI}
\PrintBackRefs{\CurrentBib}

\bibitem [\protect \citeauthoryear {%
Ropelewski%
\ \BBA {} Halpert%
}{%
Ropelewski%
\ \BBA {} Halpert%
}{%
{\protect \APACyear {1987}}%
}]{%
Ropelewski}
\APACinsertmetastar {%
Ropelewski}%
\begin{APACrefauthors}%
Ropelewski, C\BPBI F.%
\BCBT {}\ \BBA {} Halpert, M\BPBI S.%
\end{APACrefauthors}%
\unskip\
\newblock
\APACrefYearMonthDay{1987}{}{}.
\newblock
{\BBOQ}\APACrefatitle {Global and Regional Scale Precipitation Patterns Associated with the El Niño/Southern Oscillation} {Global and regional scale precipitation patterns associated with the el niño/southern oscillation}.{\BBCQ}
\newblock
\APACjournalVolNumPages{Monthly Weather Review}{115}{8}{1606 - 1626}.
\newblock
\begin{APACrefDOI} \doi{10.1175/1520-0493(1987)115<1606:GARSPP>2.0.CO;2} \end{APACrefDOI}
\PrintBackRefs{\CurrentBib}

\bibitem [\protect \citeauthoryear {%
Roy%
, Gagnon%
\BCBL {}\ \BBA {} Siingh%
}{%
Roy%
\ \protect \BOthers {.}}{%
{\protect \APACyear {2019}}%
}]{%
Roy}
\APACinsertmetastar {%
Roy}%
\begin{APACrefauthors}%
Roy, I.%
, Gagnon, A.%
\BCBL {}\ \BBA {} Siingh, D.%
\end{APACrefauthors}%
\unskip\
\newblock
\APACrefYearMonthDay{2019}{}{}.
\newblock
{\BBOQ}\APACrefatitle {Evaluating ENSO teleconnections using observations and CMIP5 models.} {Evaluating enso teleconnections using observations and cmip5 models.}{\BBCQ}
\newblock
\APACjournalVolNumPages{Theoretical and Applied Climatology}{136}{}{1085-1098}.
\newblock
\begin{APACrefDOI} \doi{https://doi.org/10.1007/s00704-018-2536-z} \end{APACrefDOI}
\PrintBackRefs{\CurrentBib}

\bibitem [\protect \citeauthoryear {%
Saha%
\ \protect \BOthers {.}}{%
Saha%
\ \protect \BOthers {.}}{%
{\protect \APACyear {2014}}%
}]{%
CFSv2}
\APACinsertmetastar {%
CFSv2}%
\begin{APACrefauthors}%
Saha, S.%
, Moorthi, S.%
, Wu, X.%
, Wang, J.%
, Nadiga, S.%
, Tripp, P.%
\BDBL {}Becker, E.%
\end{APACrefauthors}%
\unskip\
\newblock
\APACrefYearMonthDay{2014}{}{}.
\newblock
{\BBOQ}\APACrefatitle {The NCEP Climate Forecast System Version 2} {The ncep climate forecast system version 2}.{\BBCQ}
\newblock
\APACjournalVolNumPages{Journal of Climate}{27}{6}{2185--2208}.
\newblock
\begin{APACrefDOI} \doi{https://doi.org/10.1175/JCLI-D-12-00823.1} \end{APACrefDOI}
\PrintBackRefs{\CurrentBib}

\bibitem [\protect \citeauthoryear {%
Sobel%
\ \protect \BOthers {.}}{%
Sobel%
\ \protect \BOthers {.}}{%
{\protect \APACyear {2023}}%
}]{%
sobel}
\APACinsertmetastar {%
sobel}%
\begin{APACrefauthors}%
Sobel, A\BPBI H.%
, Lee, C\BHBI Y.%
, Bowen, S\BPBI G.%
, Camargo, S\BPBI J.%
, Cane, M\BPBI A.%
, Clement, A.%
\BDBL {}Tippett, M\BPBI K.%
\end{APACrefauthors}%
\unskip\
\newblock
\APACrefYearMonthDay{2023}{}{}.
\newblock
{\BBOQ}\APACrefatitle {Near-term tropical cyclone risk and coupled Earth system model biases} {Near-term tropical cyclone risk and coupled earth system model biases}.{\BBCQ}
\newblock
\APACjournalVolNumPages{Proceedings of the National Academy of Sciences}{120}{33}{e2209631120}.
\newblock
\begin{APACrefDOI} \doi{10.1073/pnas.2209631120} \end{APACrefDOI}
\PrintBackRefs{\CurrentBib}

\bibitem [\protect \citeauthoryear {%
Szunyogh%
}{%
Szunyogh%
}{%
{\protect \APACyear {2014}}%
}]{%
Szunyogh}
\APACinsertmetastar {%
Szunyogh}%
\begin{APACrefauthors}%
Szunyogh, I.%
\end{APACrefauthors}%
\unskip\
\newblock
\APACrefYear{2014}.
\newblock
\APACrefbtitle {Applicable Atmospheric Dynamics} {Applicable atmospheric dynamics}.
\newblock
\APACaddressPublisher{}{WORLD SCIENTIFIC}.
\newblock
\begin{APACrefDOI} \doi{10.1142/8047} \end{APACrefDOI}
\PrintBackRefs{\CurrentBib}

\bibitem [\protect \citeauthoryear {%
Tanaka%
, Matsumori%
, Yoshida%
\BCBL {}\ \BBA {} Aihara%
}{%
Tanaka%
\ \protect \BOthers {.}}{%
{\protect \APACyear {2022}}%
}]{%
Tanaka}
\APACinsertmetastar {%
Tanaka}%
\begin{APACrefauthors}%
Tanaka, G.%
, Matsumori, T.%
, Yoshida, H.%
\BCBL {}\ \BBA {} Aihara, K.%
\end{APACrefauthors}%
\unskip\
\newblock
\APACrefYearMonthDay{2022}{}{}.
\newblock
{\BBOQ}\APACrefatitle {Reservoir computing with diverse timescales for prediction of multiscale dynamics} {Reservoir computing with diverse timescales for prediction of multiscale dynamics}.{\BBCQ}
\newblock
\APACjournalVolNumPages{Phys. Rev. Res.}{4}{}{L032014}.
\newblock
\begin{APACrefDOI} \doi{10.1103/PhysRevResearch.4.L032014} \end{APACrefDOI}
\PrintBackRefs{\CurrentBib}

\bibitem [\protect \citeauthoryear {%
Tikhonov%
\ \BBA {} Arsenin%
}{%
Tikhonov%
\ \BBA {} Arsenin%
}{%
{\protect \APACyear {1977}}%
}]{%
Tikhonov}
\APACinsertmetastar {%
Tikhonov}%
\begin{APACrefauthors}%
Tikhonov, A\BPBI N.%
\BCBT {}\ \BBA {} Arsenin, V\BPBI Y.%
\end{APACrefauthors}%
\unskip\
\newblock
\APACrefYearMonthDay{1977}{}{}.
\newblock
{\BBOQ}\APACrefatitle {Solutions of ill-posed problems} {Solutions of ill-posed problems}.{\BBCQ}
\newblock
\APACjournalVolNumPages{Winston, Halsted Press}{}{}{}.
\PrintBackRefs{\CurrentBib}

\bibitem [\protect \citeauthoryear {%
Timmermann%
, An%
\BCBL {}\ \BBA {} Kug%
}{%
Timmermann%
\ \protect \BOthers {.}}{%
{\protect \APACyear {2018}}%
}]{%
Timmermann}
\APACinsertmetastar {%
Timmermann}%
\begin{APACrefauthors}%
Timmermann, A.%
, An, S.%
\BCBL {}\ \BBA {} Kug, J\BPBI e\BPBI a.%
\end{APACrefauthors}%
\unskip\
\newblock
\APACrefYearMonthDay{2018}{}{}.
\newblock
{\BBOQ}\APACrefatitle {El Nino-Southern Oscillation complexity} {El nino-southern oscillation complexity}.{\BBCQ}
\newblock
\APACjournalVolNumPages{Nature}{}{559}{535-545}.
\newblock
\begin{APACrefDOI} \doi{10.1038/s41586-018-0252-6} \end{APACrefDOI}
\PrintBackRefs{\CurrentBib}

\bibitem [\protect \citeauthoryear {%
Virtanen%
\ \protect \BOthers {.}}{%
Virtanen%
\ \protect \BOthers {.}}{%
{\protect \APACyear {2020}}%
}]{%
2020SciPy-NMeth}
\APACinsertmetastar {%
2020SciPy-NMeth}%
\begin{APACrefauthors}%
Virtanen, P.%
, Gommers, R.%
, Oliphant, T\BPBI E.%
, Haberland, M.%
, Reddy, T.%
, Cournapeau, D.%
\BDBL {}{SciPy 1.0 Contributors}%
\end{APACrefauthors}%
\unskip\
\newblock
\APACrefYearMonthDay{2020}{}{}.
\newblock
{\BBOQ}\APACrefatitle {{{SciPy} 1.0: Fundamental Algorithms for Scientific Computing in Python}} {{{SciPy} 1.0: Fundamental Algorithms for Scientific Computing in Python}}.{\BBCQ}
\newblock
\APACjournalVolNumPages{Nature Methods}{17}{}{261--272}.
\newblock
\begin{APACrefDOI} \doi{10.1038/s41592-019-0686-2} \end{APACrefDOI}
\PrintBackRefs{\CurrentBib}

\bibitem [\protect \citeauthoryear {%
Vitart%
}{%
Vitart%
}{%
{\protect \APACyear {2014}}%
}]{%
Vitart2014}
\APACinsertmetastar {%
Vitart2014}%
\begin{APACrefauthors}%
Vitart, F.%
\end{APACrefauthors}%
\unskip\
\newblock
\APACrefYearMonthDay{2014}{}{}.
\newblock
{\BBOQ}\APACrefatitle {Evolution of ECMWF sub-seasonal forecast skill scores} {Evolution of ecmwf sub-seasonal forecast skill scores}.{\BBCQ}
\newblock
\APACjournalVolNumPages{Quarterly Journal of the Royal Meteorological Society}{140}{683}{1889--1899}.
\newblock
\begin{APACrefURL} \url{https://doi.org/10.1002/qj.2256} \end{APACrefURL}
\newblock
\begin{APACrefDOI} \doi{https://doi.org/10.1002/qj.2256} \end{APACrefDOI}
\PrintBackRefs{\CurrentBib}

\bibitem [\protect \citeauthoryear {%
B.~Wang%
\ \protect \BOthers {.}}{%
B.~Wang%
\ \protect \BOthers {.}}{%
{\protect \APACyear {2013}}%
}]{%
Wang}
\APACinsertmetastar {%
Wang}%
\begin{APACrefauthors}%
Wang, B.%
, Liu, J.%
, Kim, H\BHBI J.%
, Webster, P\BPBI J.%
, Yim, S\BHBI Y.%
\BCBL {}\ \BBA {} Xiang, B.%
\end{APACrefauthors}%
\unskip\
\newblock
\APACrefYearMonthDay{2013}{}{}.
\newblock
{\BBOQ}\APACrefatitle {Northern Hemisphere summer monsoon intensified by mega-El Niño/southern oscillation and Atlantic multidecadal oscillation} {Northern hemisphere summer monsoon intensified by mega-el niño/southern oscillation and atlantic multidecadal oscillation}.{\BBCQ}
\newblock
\APACjournalVolNumPages{Proceedings of the National Academy of Sciences}{110}{14}{5347-5352}.
\newblock
\begin{APACrefDOI} \doi{10.1073/pnas.1219405110} \end{APACrefDOI}
\PrintBackRefs{\CurrentBib}

\bibitem [\protect \citeauthoryear {%
C.~Wang%
\ \protect \BOthers {.}}{%
C.~Wang%
\ \protect \BOthers {.}}{%
{\protect \APACyear {2024}}%
}]{%
ola}
\APACinsertmetastar {%
ola}%
\begin{APACrefauthors}%
Wang, C.%
, Pritchard, M\BPBI S.%
, Brenowitz, N.%
, Cohen, Y.%
, Bonev, B.%
, Kurth, T.%
\BDBL {}Pathak, J.%
\end{APACrefauthors}%
\unskip\
\newblock
\APACrefYearMonthDay{2024}{}{}.
\newblock
\APACrefbtitle {Coupled Ocean-Atmosphere Dynamics in a Machine Learning Earth System Model.} {Coupled ocean-atmosphere dynamics in a machine learning earth system model.}
\newblock
\begin{APACrefURL} \url{https://arxiv.org/abs/2406.08632} \end{APACrefURL}
\PrintBackRefs{\CurrentBib}

\bibitem [\protect \citeauthoryear {%
Watt-Meyer%
\ \protect \BOthers {.}}{%
Watt-Meyer%
\ \protect \BOthers {.}}{%
{\protect \APACyear {2021}}%
}]{%
Watt-Meyer2021}
\APACinsertmetastar {%
Watt-Meyer2021}%
\begin{APACrefauthors}%
Watt-Meyer, O.%
, Brenowitz, N\BPBI D.%
, Clark, S\BPBI K.%
, Henn, B.%
, Kwa, A.%
, McGibbon, J.%
\BDBL {}Bretherton, C\BPBI S.%
\end{APACrefauthors}%
\unskip\
\newblock
\APACrefYearMonthDay{2021}{}{}.
\newblock
{\BBOQ}\APACrefatitle {Correcting Weather and Climate Models by Machine Learning Nudged Historical Simulations} {Correcting weather and climate models by machine learning nudged historical simulations}.{\BBCQ}
\newblock
\APACjournalVolNumPages{Geophysical Research Letters}{48}{15}{e2021GL092555}.
\newblock
\begin{APACrefDOI} \doi{https://doi.org/10.1029/2021GL092555} \end{APACrefDOI}
\PrintBackRefs{\CurrentBib}

\bibitem [\protect \citeauthoryear {%
Weyn%
, Durran%
, Caruana%
\BCBL {}\ \BBA {} Cresswell-Clay%
}{%
Weyn%
\ \protect \BOthers {.}}{%
{\protect \APACyear {2021}}%
}]{%
weyn}
\APACinsertmetastar {%
weyn}%
\begin{APACrefauthors}%
Weyn, J\BPBI A.%
, Durran, D\BPBI R.%
, Caruana, R.%
\BCBL {}\ \BBA {} Cresswell-Clay, N.%
\end{APACrefauthors}%
\unskip\
\newblock
\APACrefYearMonthDay{2021}{}{}.
\newblock
{\BBOQ}\APACrefatitle {Sub-Seasonal Forecasting With a Large Ensemble of Deep-Learning Weather Prediction Models} {Sub-seasonal forecasting with a large ensemble of deep-learning weather prediction models}.{\BBCQ}
\newblock
\APACjournalVolNumPages{Journal of Advances in Modeling Earth Systems}{13}{7}{e2021MS002502}.
\newblock
\begin{APACrefDOI} \doi{https://doi.org/10.1029/2021MS002502} \end{APACrefDOI}
\PrintBackRefs{\CurrentBib}

\bibitem [\protect \citeauthoryear {%
Wheeler%
\ \BBA {} Kiladis%
}{%
Wheeler%
\ \BBA {} Kiladis%
}{%
{\protect \APACyear {1999}}%
}]{%
Wheeler}
\APACinsertmetastar {%
Wheeler}%
\begin{APACrefauthors}%
Wheeler, M.%
\BCBT {}\ \BBA {} Kiladis, G\BPBI N.%
\end{APACrefauthors}%
\unskip\
\newblock
\APACrefYearMonthDay{1999}{}{}.
\newblock
{\BBOQ}\APACrefatitle {Convectively Coupled Equatorial Waves: Analysis of Clouds and Temperature in the Wavenumber–Frequency Domain} {Convectively coupled equatorial waves: Analysis of clouds and temperature in the wavenumber–frequency domain}.{\BBCQ}
\newblock
\APACjournalVolNumPages{Journal of the Atmospheric Sciences}{56}{3}{374 - 399}.
\newblock
\begin{APACrefDOI} \doi{10.1175/1520-0469(1999)056<0374:CCEWAO>2.0.CO;2} \end{APACrefDOI}
\PrintBackRefs{\CurrentBib}

\bibitem [\protect \citeauthoryear {%
White%
\ \protect \BOthers {.}}{%
White%
\ \protect \BOthers {.}}{%
{\protect \APACyear {2017}}%
}]{%
White2017}
\APACinsertmetastar {%
White2017}%
\begin{APACrefauthors}%
White, C\BPBI J.%
, Carlsen, H.%
, Robertson, A\BPBI W.%
, Klein, R\BPBI J\BPBI T.%
, Lazo, J\BPBI K.%
, Kumar, A.%
\BDBL {}Zebiak, S\BPBI E.%
\end{APACrefauthors}%
\unskip\
\newblock
\APACrefYearMonthDay{2017}{}{}.
\newblock
{\BBOQ}\APACrefatitle {Potential applications of subseasonal-to-seasonal (S2S) predictions} {Potential applications of subseasonal-to-seasonal (s2s) predictions}.{\BBCQ}
\newblock
\APACjournalVolNumPages{Meteorological Applications}{24}{3}{315--325}.
\newblock
\begin{APACrefDOI} \doi{10.1002/met.1654} \end{APACrefDOI}
\PrintBackRefs{\CurrentBib}

\bibitem [\protect \citeauthoryear {%
White%
\ \protect \BOthers {.}}{%
White%
\ \protect \BOthers {.}}{%
{\protect \APACyear {2022}}%
}]{%
White2022}
\APACinsertmetastar {%
White2022}%
\begin{APACrefauthors}%
White, C\BPBI J.%
, Domeisen, D\BPBI I\BPBI V.%
, Acharya, N.%
, Adefisan, E\BPBI A.%
, Anderson, M\BPBI L.%
, Aura, S.%
\BDBL {}Wilson, R\BPBI G.%
\end{APACrefauthors}%
\unskip\
\newblock
\APACrefYearMonthDay{2022}{}{}.
\newblock
{\BBOQ}\APACrefatitle {Advances in the Application and Utility of Subseasonal-to-Seasonal Predictions} {Advances in the application and utility of subseasonal-to-seasonal predictions}.{\BBCQ}
\newblock
\APACjournalVolNumPages{Bulletin of the American Meteorological Society}{103}{6}{E1448--E1472}.
\newblock
\begin{APACrefDOI} \doi{https://doi.org/10.1175/BAMS-D-20-0224.1} \end{APACrefDOI}
\PrintBackRefs{\CurrentBib}

\bibitem [\protect \citeauthoryear {%
Wikner%
\ \protect \BOthers {.}}{%
Wikner%
\ \protect \BOthers {.}}{%
{\protect \APACyear {2020}}%
}]{%
Wikner2020}
\APACinsertmetastar {%
Wikner2020}%
\begin{APACrefauthors}%
Wikner, A.%
, Pathak, J.%
, Hunt, B.%
, Girvan, M.%
, Arcomano, T.%
, Szunyogh, I.%
\BDBL {}Ott, E.%
\end{APACrefauthors}%
\unskip\
\newblock
\APACrefYearMonthDay{2020}{}{}.
\newblock
{\BBOQ}\APACrefatitle {{Combining machine learning with knowledge-based modeling for scalable forecasting and subgrid-scale closure of large, complex, spatiotemporal systems}} {{Combining machine learning with knowledge-based modeling for scalable forecasting and subgrid-scale closure of large, complex, spatiotemporal systems}}.{\BBCQ}
\newblock
\APACjournalVolNumPages{Chaos: An Interdisciplinary Journal of Nonlinear Science}{30}{5}{053111}.
\newblock
\begin{APACrefDOI} \doi{10.1063/5.0005541} \end{APACrefDOI}
\PrintBackRefs{\CurrentBib}

\bibitem [\protect \citeauthoryear {%
You%
\ \protect \BOthers {.}}{%
You%
\ \protect \BOthers {.}}{%
{\protect \APACyear {2024}}%
}]{%
you2024}
\APACinsertmetastar {%
you2024}%
\begin{APACrefauthors}%
You, H.%
, Wang, J.%
, Wong, R\BPBI K\BPBI W.%
, Schumacher, C.%
, Saravanan, R.%
\BCBL {}\ \BBA {} Jun, M.%
\end{APACrefauthors}%
\unskip\
\newblock
\APACrefYearMonthDay{2024}{}{}.
\newblock
\APACrefbtitle {Prediction of Tropical Pacific Rain Rates with Over-parameterized Neural Networks} {Prediction of tropical pacific rain rates with over-parameterized neural networks}\ (\BVOL~3).
\newblock
\begin{APACrefDOI} \doi{10.1175/AIES-D-23-0083.1} \end{APACrefDOI}
\PrintBackRefs{\CurrentBib}

\bibitem [\protect \citeauthoryear {%
C.~Zhang%
}{%
C.~Zhang%
}{%
{\protect \APACyear {2013}}%
}]{%
Zhang2013}
\APACinsertmetastar {%
Zhang2013}%
\begin{APACrefauthors}%
Zhang, C.%
\end{APACrefauthors}%
\unskip\
\newblock
\APACrefYearMonthDay{2013}{}{}.
\newblock
{\BBOQ}\APACrefatitle {Madden--Julian Oscillation: Bridging Weather and Climate} {Madden--julian oscillation: Bridging weather and climate}.{\BBCQ}
\newblock
\APACjournalVolNumPages{Bulletin of the American Meteorological Society}{94}{12}{1849--1870}.
\newblock
\begin{APACrefDOI} \doi{https://doi.org/10.1175/BAMS-D-12-00026.1} \end{APACrefDOI}
\PrintBackRefs{\CurrentBib}

\bibitem [\protect \citeauthoryear {%
L.~Zhang%
, Kim%
, Yang%
, Hong%
\BCBL {}\ \BBA {} Zhu%
}{%
L.~Zhang%
\ \protect \BOthers {.}}{%
{\protect \APACyear {2021}}%
}]{%
zhangs2s}
\APACinsertmetastar {%
zhangs2s}%
\begin{APACrefauthors}%
Zhang, L.%
, Kim, T.%
, Yang, T.%
, Hong, Y.%
\BCBL {}\ \BBA {} Zhu, Q.%
\end{APACrefauthors}%
\unskip\
\newblock
\APACrefYearMonthDay{2021}{}{}.
\newblock
{\BBOQ}\APACrefatitle {Evaluation of Subseasonal-to-Seasonal (S2S) precipitation forecast from the North American Multi-Model ensemble phase II (NMME-2) over the contiguous U.S.} {Evaluation of subseasonal-to-seasonal (s2s) precipitation forecast from the north american multi-model ensemble phase ii (nmme-2) over the contiguous u.s.}{\BBCQ}
\newblock
\APACjournalVolNumPages{Journal of Hydrology}{603}{}{127058}.
\newblock
\begin{APACrefDOI} \doi{https://doi.org/10.1016/j.jhydrol.2021.127058} \end{APACrefDOI}
\PrintBackRefs{\CurrentBib}

\bibitem [\protect \citeauthoryear {%
Zhao%
, Zhang%
\BCBL {}\ \BBA {} Chen%
}{%
Zhao%
\ \protect \BOthers {.}}{%
{\protect \APACyear {2019}}%
}]{%
ZhaoT}
\APACinsertmetastar {%
ZhaoT}%
\begin{APACrefauthors}%
Zhao, T.%
, Zhang, Y.%
\BCBL {}\ \BBA {} Chen, X.%
\end{APACrefauthors}%
\unskip\
\newblock
\APACrefYearMonthDay{2019}{}{}.
\newblock
{\BBOQ}\APACrefatitle {Predictive performance of NMME seasonal forecasts of global precipitation: A spatial-temporal perspective} {Predictive performance of nmme seasonal forecasts of global precipitation: A spatial-temporal perspective}.{\BBCQ}
\newblock
\APACjournalVolNumPages{Journal of Hydrology}{570}{}{17-25}.
\newblock
\begin{APACrefDOI} \doi{https://doi.org/10.1016/j.jhydrol.2018.12.036} \end{APACrefDOI}
\PrintBackRefs{\CurrentBib}

\bibitem [\protect \citeauthoryear {%
Zheng%
\ \BBA {} Zhu%
}{%
Zheng%
\ \BBA {} Zhu%
}{%
{\protect \APACyear {2010}}%
}]{%
Zheng}
\APACinsertmetastar {%
Zheng}%
\begin{APACrefauthors}%
Zheng, F.%
\BCBT {}\ \BBA {} Zhu, J.%
\end{APACrefauthors}%
\unskip\
\newblock
\APACrefYearMonthDay{2010}{}{}.
\newblock
{\BBOQ}\APACrefatitle {Spring predictability barrier of ENSO events from the perspective of an ensemble prediction system} {Spring predictability barrier of enso events from the perspective of an ensemble prediction system}.{\BBCQ}
\newblock
\APACjournalVolNumPages{Global and Planetary Change}{72}{3}{108-117}.
\newblock
\begin{APACrefDOI} \doi{https://doi.org/10.1016/j.gloplacha.2010.01.021} \end{APACrefDOI}
\PrintBackRefs{\CurrentBib}

\bibitem [\protect \citeauthoryear {%
Zhou%
\ \BBA {} Zhang%
}{%
Zhou%
\ \BBA {} Zhang%
}{%
{\protect \APACyear {2023}}%
}]{%
Zhou}
\APACinsertmetastar {%
Zhou}%
\begin{APACrefauthors}%
Zhou, L.%
\BCBT {}\ \BBA {} Zhang, R\BHBI H.%
\end{APACrefauthors}%
\unskip\
\newblock
\APACrefYearMonthDay{2023}{}{}.
\newblock
{\BBOQ}\APACrefatitle {A self-attention–based neural network for three-dimensional multivariate modeling and its skillful ENSO predictions} {A self-attention–based neural network for three-dimensional multivariate modeling and its skillful enso predictions}.{\BBCQ}
\newblock
\APACjournalVolNumPages{Science Advances}{9}{10}{eadf2827}.
\newblock
\begin{APACrefDOI} \doi{10.1126/sciadv.adf2827} \end{APACrefDOI}
\PrintBackRefs{\CurrentBib}

\bibitem [\protect \citeauthoryear {%
Zuo%
, Balmaseda%
, Tietsche%
, Mogensen%
\BCBL {}\ \BBA {} Mayer%
}{%
Zuo%
\ \protect \BOthers {.}}{%
{\protect \APACyear {2019}}%
}]{%
ORAS5}
\APACinsertmetastar {%
ORAS5}%
\begin{APACrefauthors}%
Zuo, H.%
, Balmaseda, M\BPBI A.%
, Tietsche, S.%
, Mogensen, K.%
\BCBL {}\ \BBA {} Mayer, M.%
\end{APACrefauthors}%
\unskip\
\newblock
\APACrefYearMonthDay{2019}{}{}.
\newblock
{\BBOQ}\APACrefatitle {The ECMWF operational ensemble reanalysis--analysis system for ocean and sea ice: a description of the system and assessment} {The ecmwf operational ensemble reanalysis--analysis system for ocean and sea ice: a description of the system and assessment}.{\BBCQ}
\newblock
\APACjournalVolNumPages{Ocean Science}{15}{3}{779--808}.
\newblock
\begin{APACrefDOI} \doi{10.5194/os-15-779-2019} \end{APACrefDOI}
\PrintBackRefs{\CurrentBib}

\end{thebibliography}

%
%
%
%
%

\end{document}